\documentclass[11pt]{article}
\usepackage[textwidth=15.2cm,textheight=22cm]{geometry}
\usepackage{amsmath,amssymb}
\usepackage{latexsym}
\usepackage{graphicx}
\usepackage{bbm}
\usepackage{bm}
\usepackage{cite}

\allowdisplaybreaks[1]

\numberwithin{equation}{section} 
                                                                               
\newcommand{\er}{{\rm e}}
\newcommand{\dr}{{\rm d}}
\newcommand{\Tr}{{\rm Tr}}

\newcommand{\R}{\mathbbm{R}}

\newcommand{\del}{\partial}
\newcommand{\be}{\begin{equation}}
\newcommand{\ee}{\end{equation}}
\newcommand{\ba}{\begin{eqnarray}}
\newcommand{\ea}{\end{eqnarray}}
\newcommand{\bdm}{\begin{displaymath}}
\newcommand{\edm}{\end{displaymath}}
\newcommand{\ra}{\rangle}
\newcommand{\la}{\langle}

\newcommand\fr[1]{\frac{1}{#1}}

\newcommand{\Dsm}{\,{\raisebox{0.5pt}{$/$} \hspace{-7.5pt} D}}

\def\b{\beta}
\def\a{\alpha}
\def\g{\gamma}

\def\veps{\varepsilon}

\def\d{\delta}

\def\v{\varphi}
\newcommand{\half}{\frac{1}{2}}
\newcommand{\nn}{\nonumber}

\newcommand{\hepth}[1]{{\tt hep-th/#1}}
\newcommand{\hepph}[1]{{\tt hep-ph/#1}}
\newcommand{\calN}{{\mathcal N}}

\newcommand{\calS}{{\mathcal S}}

\DeclareMathAlphabet{\mathpzc}{OT1}{pzc}{m}{it}
\newcommand\hsp[1]{\hspace*{#1 cm}}
\newcommand\vsp[1]{\vspace*{#1 cm}}
\newcommand{\ndt}{\noindent}
\newcommand{\nbar}[1]{\overline{#1}}

\def\bea{\begin{eqnarray}}
\def\eea{\end{eqnarray}}
\def\beas{\begin{eqnarray*}}
\def\eeas{\end{eqnarray*}}
\def\sla{\raise.15ex\hbox{$/$}\kern-.57em}

\def\parm{{\partial}_{-}}

\newcommand{\rd}{{\rm d}}

\newcommand{\vcd}{\varphi}

\begin{document}

\begin{titlepage}
\begin{flushright}    
{\small AEI-2006-067 \\
TCDMATH 06--12}
\end{flushright}
\vskip 1cm

\centerline{\LARGE{\bf {Proof of all-order finiteness for}}}
\vskip .3cm
\centerline{\LARGE{\bf {planar $\b$-deformed Yang--Mills}}}

\vskip 1.5cm

\centerline{Sudarshan Ananth$^\dagger$, Stefano Kovacs$^{\,*\,\dagger}$ 
and Hidehiko Shimada$^\dagger$}

\vskip .5cm

\centerline{{\it {$^\dagger$ Max-Planck-Institut f\"{u}r 
Gravitationsphysik}}}
\centerline{{\it {$\;$ Albert-Einstein-Institut, Potsdam, Germany}}}

\vskip 0.5cm

\centerline{\em $^*$ School of Mathematics}
\centerline{\em $\;$ Trinity College, Dublin, Ireland}

\vskip 1.5cm

\centerline{\bf {Abstract}}
\vskip .5cm

\noindent 
We study a marginal deformation of $\calN=4$ Yang--Mills, with a real
deformation parameter $\beta$. This $\beta$-deformed model has only
$\calN=1$ supersymmetry and a U(1)$\times$U(1) flavor symmetry. The
introduction of a new superspace $\star$-product allows us to
formulate the theory in $\calN=4$ light-cone superspace, despite the
fact that it has only $\calN=1$ supersymmetry.  We show that this
deformed theory is conformally invariant, in the planar approximation,
by proving that its Green functions are ultra-violet finite to all
orders in perturbation theory.

\vfill

\end{titlepage}

\section {Introduction}
\label{intro}

One of the most striking consequences of supersymmetry is an
improvement in the ultra-violet behavior of quantum field theories.
This feature provides one of the main motivations for the study of
supersymmetric models in connection with the resolution of the
hierarchy problem. 

The maximally supersymmetric $\mathcal N=4$ Yang--Mills~\cite{N4}
theory is particularly remarkable in that it is ultra-violet
finite~\cite{finN4,SM,BLN2}, therefore representing an example of a
four-dimensional quantum field theory which is conformally invariant
at the quantum level. The discovery of the special properties of the
$\calN=4$ supersymmetric Yang--Mills (SYM) theory has led to extensive
investigations aimed at identifying field theories with the same
finiteness properties, but a smaller amount of
supersymmetry~\cite{oldfinite}. A special class of such theories are
those obtained as exactly marginal deformations of $\calN=4$
SYM. These were classified in~\cite{LS}, where it was argued that
there exists a two (complex) parameter family of such deformations.  In
this paper we shall provide a proof of finiteness, to all orders in
planar perturbation theory, for a particular model in this class,
which involves a single real deformation parameter.

The existence of families of marginal deformations of $\calN=4$ SYM
has interesting consequences in the context of the AdS/CFT
correspondence~\cite{adscft}. In the framework of the AdS/CFT duality
the conformal symmetry of the boundary field theory is associated with
the isometry group of the dual anti-de Sitter background. Therefore
marginal deformations of $\calN=4$ SYM should correspond to
deformations of the AdS$_5\times S^5$ background which preserve the
SO(4,2) group of isometries of the AdS$_5$ factor.  This
correspondence was first considered in~\cite{bl} and further studied
in~\cite{AKY}, where, by expanding the supergravity equations around
the AdS$_5\times S^5$ solution, it was found that the dimension of the
space of solutions preserving $\calN=1$ supersymmetry as well as the
SO(4,2) isometry group is the same as that of the space of marginal
deformations of $\calN=4$ SYM~\cite{LS}. An exact supergravity
solution dual to a $\calN=1$ superconformal field theory with a
U(1)$\times$U(1) flavor symmetry was constructed in~\cite{LM}. The
theory dual to the supergravity background of~\cite{LM} belongs to the
class of marginal deformations of $\calN=4$ SYM referred to as
$\b$-deformations, which are characterized, in $\calN=1$ superspace,
by a superpotential of the form
\be
{\rm W}=\int \dr^4x \left[ \int\dr^2\theta \: g\,h
\,\Tr\left(\er^{i\pi\b} \Phi^1\Phi^2\Phi^3 - \er^{-i\pi\b}
\Phi^1\Phi^3\Phi^2\right) + {\rm h.c.} \right] ,
\label{genbdef}
\ee
where $g$ is the standard Yang--Mills coupling and $h$ and $\b$ are
two complex deformation parameters. In (\ref{genbdef})
$\Phi^1,\Phi^2$ and $\Phi^3$ are three $\calN=1$ chiral
superfields. Following~\cite{LM} the  perturbative properties of
theories in this class have been extensively
studied~\cite{fg,rss,milan,vk}.

In this paper we focus on the special case in which the superpotential
(\ref{genbdef}) has $h=1$ and $\b\in\R$. We prove that this
deformation of $\calN=4$ SYM is conformally invariant in the planar
limit by showing that all the Green functions in the theory are
finite. In order to prove this result to all orders in perturbation
theory we shall realize the deformation by the introduction of new
star-products acting in superspace. This will allow us to formulate
the theory in $\calN=4$ light-cone superspace and to use the same
arguments utilized in~\cite{SM,BLN2} to prove the ultra-violet
finiteness of $\calN=4$ SYM. For an introduction to light-cone
superspace, we refer the reader to~\cite{BLN1,BBB2,SA1}. The
super-Poincar\'e and superconformal algebras were presented in
light-cone superspace in references~\cite{BBB2,SA1,ABR1,ABKR}. The
truncation of supersymmetry and superfields in this context was
discussed in~\cite{BT,SA2}.

The formulation of the deformed Yang--Mills theory using star-products
presents analogies with the construction of non-commutative field
theories~\cite{noncomm}. However, the theory that we study is an
ordinary gauge theory and these analogies are purely formal. In
particular, we stress that the $\b$-deformed Yang--Mills theory
retains a non-trivial dependence on the deformation parameter in the
planar limit.

As will be explicitly shown, the proof of~\cite{SM,BLN2} can only be
applied to the deformed model in the planar approximation. The
analysis that we present is valid for arbitrary gauge group, but the
case of SU($N$) is particularly interesting, since in this case the
theory is expected to be dual to the background of~\cite{LM}. The
all-order finiteness of the $\b$-deformed Yang--Mills theory in the
planar limit suggests that the inclusion of string tree-level
corrections should not break the SO(4,2) isometries of the
supergravity solution of~\cite{LM}. Our formalism does not allow us to
reach any conclusion about the finiteness of the theory for finite $N$
and thus about the effect of string loop corrections on the dual
background.

This paper is organized as follows. In sections \ref{beta} and
\ref{lcf} we show how to formulate the $\b$-deformed theory in
$\calN=4$ light-cone superspace. Section \ref{proof} presents the
proof of finiteness to all orders in perturbation theory following the
$\calN=4$ analysis of~\cite{SM,BLN2}. Various appendices discuss
aspects of the superspace calculations which are affected by the
introduction of star-products.

\section{$\b$-deformed Yang--Mills}
\label{beta}

The $\b$-deformed Yang--Mills theory has the same field content as
$\calN=4$ Yang--Mills. In terms of $\calN=1$ multiplets, it consists of
three chiral multiplets, $(\phi^I,\lambda^I_\a)$, $I=1,2,3$ in the
adjoint representation of the gauge group and a vector multiplet,
$(A_\mu,\lambda^4_\a)$.  The deformed theory has only $\calN=1$
supersymmetry and, in addition to the U(1) R-symmetry, it has a
U(1)$\times$U(1) flavor symmetry. The three complex scalars,
$\phi^I$, and the three Weyl fermions, $\lambda^I_\a$, are charged
under the U(1)$\times$U(1) symmetry while the fields in the vector
multiplet are not. The deformation is realized simply by replacing
all the commutators of charged fields in the $\mathcal N=4$ Yang--Mills
action by $*$-commutators~\cite{LM}. The $*$-commutator is defined by
\bea
{[f,g]_*}=f*g-g*f\, ,
\eea
where 
\be
f*g=\er^{i\pi\b(Q^1_fQ^2_g-Q^2_fQ^1_g)}fg\, .
\ee
The charges, $Q^1$ and $Q^2$, of the various fields with respect to the
two U(1) flavor symmetries are read off from table \ref{flavortable}.

\begin{table}[!htb]
\begin{center}
\begin{tabular}{|c|c|c|}
\hline
Field \raisebox{-3pt}{\rule{0pt}{14pt}} & U(1)$_1$ & U(1)$_2$  \\
\hline
$\phi^1,\lambda^1$ \raisebox{-3pt}{\rule{0pt}{14pt}} & $\;\;\,0$ & $-1$  \\
\hline
$\phi^2,\lambda^2$ \raisebox{-3pt}{\rule{0pt}{14pt}} & $+1$ & $+1$ \\
\hline
$\phi^3,\lambda^3$ \raisebox{-3pt}{\rule{0pt}{14pt}} & $-1$ & $\;\;\,0$ \\
\hline
\end{tabular}
\end{center}
\caption{Flavor charges of the fields in $\b$-deformed $\calN=4$ 
Yang--Mills.}
\label{flavortable}
\end{table}

\ndt
The $\b$-deformation breaks the $\calN=4$ supersymmetry
down to $\calN=1$. The resulting deformed action reads
\ba
\calS &\!\!=\!\!& \int \dr^4x\,\Tr \left\{
\frac{1}{2}F^{\mu\nu}F_{\mu\nu}+2(D^\mu{\bar \phi}_I ) (D_\mu
\phi^I) -2i\,\lambda^I \Dsm \bar\lambda_I 
-2i\,\lambda^4 \Dsm\bar\lambda_4 \right. \nn \\
&& \hsp{1.2} - 2\sqrt{2}\,g \left[\left( 
\epsilon_{IJK}[\lambda^I,\lambda^J]_* \phi^K +\epsilon^{IJK}
[{\bar\lambda}_I,{\bar \lambda}_J]_*{\bar \phi}_K\right)-
([\lambda^4,\lambda^I]\bar\phi_I + [{\bar \lambda}_4,
{\bar \lambda}_I]\phi^I) \right] \nn \\
&& \hsp{1.15}\left. + g^2\left(\half [\phi^I,{\bar \phi}_I]
[\phi^J,{\bar \phi}_J] - [\phi^I,\phi^J]_*[{\bar \phi}_I,
{\bar \phi}_J]_* \right) \right\} \, .
\label{Ldef} 
\ea
In this paper, we restrict ourselves to the case where the deformation
parameter $\b$ is real. This ensures that the action as written in
(\ref {Ldef}) is real.

In the following we will formulate the theory in $\calN=4$ light-cone
superspace and so it is convenient to use a manifestly SU(4)
notation. We introduce scalar fields, $\v^{mn}$, $m,n=1,2,3,4$,
satisfying
\bea
\v^{mn}=-\v^{nm}\, ,\qquad {{\bar \v}_{mn}}=(\v^{mn})^\dagger
=\frac{1}{2}\,{\epsilon_{mnpq}}\,{\vcd^{pq}}\ 
\eea
and related to the $\phi^I$'s by
\bea
\phi^I=2\,\v^{I4}\, ,\qquad {\bar \phi}_I=
\epsilon_{IJK4}\,\v^{JK}\, ,\qquad I,J,K=1,2,3 .
\eea
The fermion fields are combined into
\be
\lambda^m_\a=(\lambda^I_\a,\lambda^4_\a)\, .
\ee

\section{The light-cone formalism}
\label{lcf}

We now proceed to formulate the $\b$-deformed theory of the
previous section in the light-cone gauge. With the space-time metric
$(-,+,+,+)$, the light-cone coordinates and their derivatives are
\begin{align}
{x^{\pm}}=\frac{1}{\sqrt 2}({x^0}{\pm}{x^3})\ &,\quad 
x=\frac{1}{\sqrt 2}({x^1}+i{x^2})\, ,\quad 
{\bar x}=\frac{1}{\sqrt 2}({x^1}-i{x^2})\, , \nn \\
{\partial_{\pm}}=\frac{1}{\sqrt 2}\left(\frac{\del}{\del x^0}\pm
\frac{\del}{\del x^3}\right)\, , & \quad  {\bar\partial}=
\frac{1}{\sqrt 2}\left(\frac{\del}{\del x^1}- i\frac{\del}{\del x^2}
\right)\, , \quad {\partial} =\frac{1}{\sqrt 2}\left(
\frac{\del}{\del x^1} + i\frac{\del}{\del x^2}\right)\, .
\end{align}

\subsection{The light-cone component description}
\label{lcc}

The choice of light-cone gauge involves eliminating the unphysical
degrees of freedom. In the case of the gauge field, which splits into
light-cone components, $A_+,A_-,A$ and $\bar A$ this involves setting
\be
A_-=0 
\ee
and using the equations of motion to solve for $A_+$. On the
light-cone, the fermions split up as
\bea
\lambda^m_\a\rightarrow\,(\chi^{m(+)},\chi^{m(-)})\, .
\eea
The equations of motion allow us to eliminate $\chi^{m(+)}$. For
simplicity of notation, we drop the $\scriptstyle (-)$ index from the
physical component, $\chi^{m(-)}$. For details regarding the
derivation of the light-cone action in the non-deformed case, we refer
the reader to~\cite{BLN1}.

The introduction of $*$-commutators into the action does not affect
the validity of the light-cone procedure. We stress that this point is
non-trivial because of the non-linearity of the equations of
motion. In appendix \ref{starprop}, we list the properties of the
$*$-product that ensure that this procedure remains unaffected. We
find that the light-cone component action describing $\b$-deformed
Yang--Mills is
\ba
\label{lca}
S &\!\!=\!\!& \int \dr^4x \, \Tr \left\{2 \bar A\Box A
+ \half\, \bar\v_{mn}\Box\v^{mn}-\frac{2i}{\sqrt{2}}\,\bar\chi_m
\frac{\Box}{\del_-} \chi^m\right. \nn \\
&& + g\left[ 4i\,\frac{\bar\del}{\del_-}A [\del_-
\bar A,A]_*  +i\, \frac{\bar\del}{\del_-} A
[\del_-\bar\v_{mn},\v^{mn}]_* -i \,A[\bar\del\bar\v_{mn},
\v^{mn}]_* \right. \nn \\
&&\left. \hsp{0.7}-2\sqrt{2}\,\frac{\bar\del}{\del_-} 
A[\bar\chi_m,\chi^m]_*+2\sqrt{2}\,A[\chi^m,\frac{\bar\del}{\del_-}
\bar\chi_m]_* -2\sqrt{2}\frac{\bar\del}{\del_-}\bar\chi_m
[\bar\chi_n,\v^{mn}]_* + {\rm h.c.} \right] \nn \\
&& + g^2 \left[ 4\fr{\del_-} [\del_- A,\bar A]_*\fr{\del_-}
[\del_-\bar A,A]_* +[\v^{mn},A]_*[\bar\v_{mn},\bar A]_* \right. \nn \\
&& \hsp{0.9} + \fr{\del_-}[\del_-\bar A,A]_*\fr{\del_-}[\del_-\bar\v_{mn},
\v^{mn}]_* + \fr{\del_-} [\del_- A,\bar A]_* \fr{\del_-}
[\del_-\bar\v_{mn},\v^{mn}]_* \nn \\
&& \hsp{0.9}+\fr{8} [\v^{mn},\v^{pq}]_*[\bar\v_{mn},\bar\v_{pq}]_*
+\fr{4}\,\fr{\del_-}[\del_-\bar\v_{mn},\v^{mn}]_*\fr{\del_-}
[\del_-\bar\v_{pq},\v^{pq}]_* \nn \\
&& \hsp{0.9} -i2\sqrt{2}\,\fr{\del_-}[\bar\chi_m,\bar A]_*
[A,\chi^m]_*+i2\sqrt{2}\,\fr{\del_-}[\chi^m,A]_*
[\bar\v_{mn},\chi^n]_* \nn \\
&& \hsp{0.9} +i2\sqrt{2}\,\fr{\del_-}[\bar\chi_m,\bar A]_*
[\v^{mn},\bar\chi_n]_* + i2\sqrt{2}\,\fr{\del_-}[\bar\chi_m,\v^{mn}]_*
[\bar\v_{np},\chi^p]_* \nn \\
&& \hsp{0.9} +i2\sqrt{2}\,\fr{\del_-}[\del_-A,\bar A]_*\fr{\del_-}
[\bar\chi_m,\chi^m]_* +i2\sqrt{2}\,\fr{\del_-}[\del_-\bar A,A]_*
\fr{\del_-}[\bar\chi_m,\chi^m]_* \nn \\
&& \left. \left.\hsp{0.9} +i\sqrt{2}\, \fr{\del_-}
[\del_-\bar\v_{mn},\v^{mn}]_* \fr{\del_-}[\bar\chi_p,\chi^p]_* 
-2 \fr{\del_-}[\bar\chi_m,\chi^m]_*\fr{\del_-}[\bar\chi_n,\chi^n]_* 
\right]\right\}\ ,
\ea
where we use for the $\frac{1}{\parm}$ operator the prescription given
in~\cite{SM}. As in the covariant case, this light-cone component
action is obtained by replacing all the commutators in the $\calN=4$
light-cone action~\cite{BLN1} by $*$-commutators. Notice, however,
that many of the $*$-commutators are actually ordinary commutators
because of charge neutrality.

\subsection{Light-cone superspace formalism}
\label{lcss}

The $\b$-deformed theory has $\mathcal N=1$ supersymmetry. Despite
this we will show that the theory can be formulated in $\calN=4$
light-cone superspace thanks to the fact that its field content is
identical to that of $\calN=4$ Yang--Mills. This is achieved by
introducing a new star product in superspace which implements the
effects of the deformation introduced into the component action. The
$\calN=4$ light-cone superspace~\cite{BLN1,BBB2,SA1,ABR1,ABKR,BT,SA2}
is made up of four bosonic coordinates, $x^+,x^-,x,\bar x$, and eight
fermionic coordinates~\footnote{These Grassmann coordinates do not
carry spinor indices.}, $\theta^m,{\bar \theta}_m$, $m=1,2,3,4$. These
will be collectively denoted by $z=(x^+,x^-,x,{\bar x},\theta^m,{\bar
\theta}_m)$. 

All the degrees of freedom of the deformed theory are described by a
single scalar superfield~\cite{BLN1}. This superfield is defined by
the constraints
\be
\label{chir}
d^m\,\Phi=0\,,\qquad {\bar d}_n\,{\bar \Phi}=0\,, 
\ee
as well as the ``inside-out constraints''
\be
\label{io}
{\bar d}_m{\bar d}_n\Phi=\frac{1}{2}\epsilon_{mnpq}d^pd^q{\bar \Phi}\, ,
\ee
where $\bar \Phi$ is the complex conjugate of $\Phi$. The superspace
chiral derivatives in the above expressions are
\be
{d^{\,m}}=-\frac{\partial}{\partial {\bar \theta}_m}\,+\,
\frac{i}{\sqrt 2}\,{\theta^m}\,
{\partial_-}\,,\qquad{{\bar d}_{\,n}}=\frac{\partial}{\partial \theta^n}\,
-\,\frac{i}{\sqrt 2}\,{{\bar \theta}_n}\,{\partial_-}
\ee
and obey
\be
\{d^m,{\bar d}_n\}=i{\sqrt 2}\,\delta^m_n{\parm}\, .
\ee
The superfield satisfying the constraints (\ref {chir}) and (\ref
{io}) is~\cite{BLN1}
\begin{align}
\label{superf}
&\Phi(x,\theta,\bar\theta)=-\,\frac{1}{\parm}A(y)-\frac{i}{\parm}\theta^m
{\bar \chi}_m(y)+\frac{i}{\sqrt 2}\theta^m\theta^n
{\nbar \vcd}_{mn}(y) \nn \\
&+\frac{\sqrt 2}{6}\theta^m\theta^n\theta^p\epsilon_{mnpq}
\chi^q(y)-\frac{1}{12}\theta^m\theta^n\theta^p\theta^q
\epsilon_{mnpq}\parm\,{\bar A}(y)\ , 
\end{align}
where $y=\,(\,x,\,{\bar
x},\,{x^+},\,y^-\,\equiv\,{x^-}-\,\frac{i}{\sqrt 2}\,{\theta^m}\,{\bar
\theta}_m\,)$ is the chiral coordinate and the r. h. s. of
(\ref{superf}) is understood as a power expansion around $x^-$.

\subsubsection{The superspace $\star$-product}
\label{star}

The deformation in the component action is realized using the
$*$-commutator. In order to formulate the deformed theory in
superspace, we introduce a new superspace $\star$-product whose effect
on superfields mimics the action of the $*$-product on component
fields.

Star-products differ from ordinary products by phase factors. In the
component formulation, these phase factors were associated with the
charges carried by the component fields~\cite{LM}. In superspace, we
will think of these phase factors as coming from the $\theta$'s
instead. This is possible because each component field in a superfield
is accompanied by a unique combination of $\theta$'s. Based on table
\ref{flavortable} we assign to the $\theta$ variables (the charges of
the ${\bar \theta}$'s are opposite to those of the $\theta$'s) the
U(1)$\times$U(1) charges in table~\ref{thetacharges}.

\begin{table}[!htb]
\begin{center}
\begin{tabular}{|c|c|c|}
\hline
Variable \raisebox{-3pt}{\rule{0pt}{14pt}} & U(1)$_1$ & U(1)$_2$  \\
\hline
$\theta^1$ \raisebox{-3pt}{\rule{0pt}{14pt}} & $\;\;\,0$ & $-1$  \\
\hline
$\theta^2$ \raisebox{-3pt}{\rule{0pt}{14pt}} & $+1$ & $+1$ \\
\hline
$\theta^3$ \raisebox{-3pt}{\rule{0pt}{14pt}} & $-1$ & $\;\;\,0$ \\
\hline
$\theta^4$ \raisebox{-3pt}{\rule{0pt}{14pt}} & $\;\;\,0$ & $\;\;\,0$ \\
\hline
\end{tabular}
\end{center}
\caption{$\theta$ charges under the flavor symmetry.}
\label{thetacharges}
\end{table}

\ndt
With this assignment the superspace $\star$-product is realized in
terms of operators which count the number of $\theta$'s and $\bar
\theta$'s.  

We define the $\star$-product of two superfields, $F$ and $G$, by
\be
\label{sstar}
F \star G = F\,\er^{i\pi{\b}({\overleftarrow Q}^1_F
{\overrightarrow Q}^2_G-{\overleftarrow Q}^2_F
{\overrightarrow Q}^1_G)}\,G\,,
\ee
where the charges are the operators
\bea
\label{charges}
&&\hsp{-2}{\overrightarrow Q}^1=\theta^2\frac{{\overrightarrow
{\partial}}}{\partial\theta^2}-\theta^3\frac{{\overrightarrow
{\partial}}}{\partial\theta^3}-{\bar \theta}_2\frac{{\overrightarrow
{\partial}}}{\partial{\bar \theta}_2}+{\bar \theta}_3
\frac{{\overrightarrow {\partial}}}{\partial{\bar \theta}_3}\, , \nn \\
&&\hsp{-2}{\overleftarrow Q}^1={\frac{{\overleftarrow
{\partial}}}{\partial\theta^2}}\theta^2-\frac{{\overleftarrow
{\partial}}}{\partial\theta^3}\theta^3-\frac{{\overleftarrow
{\partial}}}{\partial{\bar \theta}_2}{\bar
\theta}_2+\frac{{\overleftarrow {\partial}}}{\partial{\bar
\theta}_3}{\bar \theta}_3\, , \nn \\ 
&&\hsp{-2}{\overrightarrow Q}^2=\theta^2\frac{{\overrightarrow
{\partial}}}{\partial\theta^2}-\theta^1\frac{{\overrightarrow
{\partial}}}{\partial\theta^1}-{\bar \theta}_2\frac{{\overrightarrow
{\partial}}}{\partial{\bar \theta}_2}+{\bar
\theta}_1\frac{{\overrightarrow {\partial}}}
{\partial{\bar \theta}_1}\, ,\nn \\ 
&&\hsp{-2}{\overleftarrow Q}^2={\frac{{\overleftarrow
{\partial}}}{\partial\theta^2}}\theta^2-\frac{{\overleftarrow
{\partial}}}{\partial\theta^1}\theta^1-\frac{{\overleftarrow
{\partial}}}{\partial{\bar \theta}_2}{\bar
\theta}_2+\frac{{\overleftarrow {\partial}}}{\partial{\bar
\theta}_1}{\bar \theta}_1\, .
\eea
The various terms in the $\theta$-expansion of a superfield, as in
(\ref {superf}), have definite U(1)$\times$U(1) charges. Therefore,
after substituting the $\theta$-expansion in the $\star$-product of
two superfields, the operators in (\ref{charges}) acting on each term
in the sum produce definite phases according to the charge assignments
in table \ref{thetacharges}. This is useful as it makes the
manipulation  of $\star$-products in superspace expressions much
simpler. Notice that, although in the $\calN=4$ light-cone superspace
formulation the phase factors introduced by the $\star$-products are
associated with the $\theta^m$ and $\bar\theta_m$ fermionic
coordinates, the U(1)$\times$U(1) symmetry is an ordinary flavor
symmetry. This is to be distinguished from the U(1)$_R$ symmetry which
is the standard $\calN=1$ R-symmetry and acts both on the $\theta$'s
and $\bar\theta$'s and on the component fields.

\subsubsection{The action}
\label{act}

The light-cone superspace action for $\b$-deformed Yang--Mills
is~\footnote{This is an ordinary ``commutative" field theory although
the definition in (\ref {sstar}) suggests the introduction of
non-commutativity.}
\bea
\label{ans}
S&\!\!\!=\!\!\!& 72\int\dr^4x \int\dr^4\theta\,
{\rd}^4{\bar \theta}\;\Tr\left\{-2\,{\bar \Phi}\,
\frac{\Box}{\parm^2}\,\Phi
+i\,\frac{8}{3}\,g \left(\frac{1}{\parm}{\bar \Phi}\,
[\Phi,{\bar \partial}\Phi]_\star\,+\,\frac{1}{\parm}\,\Phi\,
[{\bar \Phi},\partial{\bar \Phi}]_\star\right) \right. \nn \\
&& \hsp{1}\left. +\,2\,g^2\left(\frac{1}{\parm}[\Phi,\parm\Phi]_\star\,
\frac{1}{\parm}[{\bar \Phi},\parm{\bar \Phi}]_\star\,
+\fr{2}[\Phi,{\bar \Phi}]_\star\,[\Phi,{\bar \Phi}]_\star \right)\right\} \, .
\eea
Expanding the $\star$-commutators and performing the Grassmann
integrations reproduces exactly (\ref {lca}). This justifies our
definition of the superspace $\star$-product. This deformed theory is
formulated in a manifestly $\calN=4$ superspace, but because of the
presence of the $\star$-products only one supersymmetry remains
unbroken. We stress that this action can be obtained from the
light-cone superspace action of~\cite{BLN1} by just
replacing ordinary commutators by $\star$-commutators.

The action in (\ref {ans}) can be expressed purely in terms of the
chiral superfield. This is possible using the inside-out constraint in
(\ref {io}) which implies that
\be
{\bar \Phi}=\fr{48}\frac{\bar d^4}{\parm^2}\Phi\, ,
\ee
where
\be
\bar d^4 = \veps^{mnpq}\,\bar d_m\bar d_n\bar d_p\bar d_q \, .
\label{dbar4def}
\ee 
Unless otherwise indicated this notation will be used in all the
following formulae. Similarly we will use
\be
d^4=\veps_{mnpq}\,d^md^nd^pd^q \, .
\label{d4def}
\ee
The rules for partially integrating chiral derivatives
in the action are modified due to the presence of the
$\star$-products. These modified manipulations of the chiral
derivatives are explained in appendix \ref {ssstar}.

From the kinetic term in the action, we read off the
propagator (in this equation, we make explicit the matrix indices on
$\Phi$)
\bea
\la{(\Phi)}^u_{\;\;v}(z_1)\,{(\Phi)}^r_{\;\;s}(z_2)\ra
=\la\Phi^a(z_1)\,{(T^a)}^u_{\;\;v}\,\Phi^b(z_2)\,{(T^b)}^r_{\;\;s}\ra
=\Delta^{u\;r}_{\;\;v\;s}\,(z_1-z_2)\, ,
\eea
where $z=(x^+,x^-,x,{\bar x},\theta,{\bar \theta})$ and $T^a,T^b$ are
representation matrices for the Lie algebra. The corresponding
momentum-space propagator reads
\bea
\label{prop}
\Delta^{u\,r}_{\;\;v\,s}(k,\theta_{(1)},{\bar \theta}_{(1)},\theta_{(2)},
{\bar \theta}_{(2)})\,=\,t^{u\,r}_{\;\;v\,s}\,\fr{k_\mu^2}\,d_{(1)}^4\,
\delta^8(\theta_{(1)}-\theta_{(2)})\, ,
\eea
where $\theta_{(1)}$ and $\theta_{(2)}$ denote the fermionic
coordinates at points $z_1$ and $z_2$ respectively and
$t^{u\,r}_{\;\;v\,s}$ is a tensor whose precise structure depends on
the choice of gauge group. In our calculations, we use matrix notation
and this tensor will not appear explicitly. The fermionic
$\delta$-function is
\be
\delta^8(\theta_{(1)}-\theta_{(2)})=\left(\theta_{(1)}-\theta_{(2)}
\right)^4 \left(\bar\theta_{(1)}-\bar\theta_{(2)}\right)^4 \, .
\label{fdelta}
\ee
In appendix \ref{wick}, we further streamline our
notation and display a sample Wick contraction.

\section {Proof of finiteness}
\label{proof}

In this section we explicitly prove that all light-cone superspace
Green functions in the $\b$-deformed theory are finite in the
planar limit. Having realized the $\b$-deformation in the manner
described in the previous section the proof of finiteness
of~\cite{BLN2} can be repeated step by step in the present case.

The general philosophy behind our approach is as follows. 
\begin{itemize}
\item The superficial degree of divergence of all planar supergraphs
can be shown to be zero using a version of the power counting methods
of~\cite{GRS}, adapted to our formalism. This result assumes that all
momenta in a supergraph contribute to the loop integral and provides a
preliminary estimate.
\item We then distinguish between internal and external momenta and
focus on vertices attached to external legs. Using manipulations of
the chiral derivatives we show that the superficial degree of
divergence can be reduced to a negative value.
\item The above analysis applies to the entire supergraph. The same
analysis can be applied to prove that all subgraphs also have negative
superficial degree of divergence.
\item Having shown that all supergraphs and their subgraphs in the
planar approximation have negative superficial degree of divergence,
finiteness of all Green functions follows from Weinberg's
theorem~\cite{SW}.
\end{itemize}

\subsection{Supergraph power counting}
\label{sgpc}

In this subsection we explain how the superficial degree of divergence is
estimated. A general procedure for determining the degree of
divergence of diagrams in superspace was developed
in~\cite{GRS}. These power counting rules were applied to $\calN=4$
Yang--Mills in light-cone superspace in~\cite{BLN2} to show that all
supergraphs are at most logarithmically divergent if all momenta
contribute to the loop integrals. This result remains valid in the
$\b$-deformed theory at the planar level, whereas the same analysis
provides a less stringent bound on the divergence of non-planar
diagrams.

The first step in the analysis of the degree of divergence is to
perform the fermionic integrals using the $\delta$-functions in the
propagator (\ref{prop}). This procedure is easily repeated for all the
$\theta$-integrals within a given loop until $\theta$ integrations at
only two superspace points remain. The last two integrals are
evaluated using the formula~\cite{BLN2,GRS}
\be
\label{grsN4}
\d^8(\theta_{(1)}-\theta_{(2)})\,d_{(1)}^4\,\bar d_{(1)}^4\,
\d^8(\theta_{(1)}-\theta_{(2)})=\d^8(\theta_{(1)}-\theta_{(2)})\, ,
\ee
which shows that when acting with the operator $d^4_{(1)} \bar
d^4_{(1)}$ on the second $\delta$-function only the fermionic
derivatives contribute. This implies that (\ref{grsN4}), which could
have potentially contributed four powers of momentum to the loop
integral, instead has a null contribution. Note that any other
combination of the chiral derivatives, involving less than four $d$'s
and four $\bar d$'s, is zero because it necessarily involves factors
of $(\theta_{(1)}-\theta_{(1)})$. This result combined with the usual
power counting rules, implies that the superficial degree of divergence
of a generic supergraph in $\calN=4$ Yang--Mills is zero~\cite{SM,BLN2}.

This analysis is in general affected by the $\b$-deformation. The
effect of the modification described in section~\ref{act} is to
introduce $\star$-products into supergraphs. In the case of planar
supergraphs the step-wise fermionic integration leads to expressions
of the form 
\be
\d^8(\theta_{(1)}-\theta_{(2)})\left[A(\theta_{(1)})
\star_{(1)}d_{(1)}^4\bar d_{(1)}^4 \d^8(\theta_{(1)}-\theta_{(2)})
\star_{(2)}B(\theta_{(2)})\right], 
\label{stardd}
\ee
where $A(\theta_1)$ and $B(\theta_2)$ are arbitrary superfields and
$\star_{(1)}$, $\star_{(2)}$ act at superspace points $1$, $2$. The
presence of $\star$-products introduces phases implying that the
expansion of the expression in brackets contains factors such as
\be
\label{factor}
(\er^{i\pi{p_1}}\theta_{(1)}-\er^{i\pi{p_2}}\theta_{(1)})\, .
\ee
In planar diagrams, charge conservation ensures that $p_1=p_2$, so we
have a formula analogous to (\ref {grsN4})
\be
\label{grsN42}
\d^8(\theta_{(1)}-\theta_{(2)})\!\left[A(\theta_{(1)})\star_{(1)} 
d_{(1)}^4\bar d_{(1)}^4 \d^8(\theta_{(1)}-\theta_{(2)})
\star_{(2)}B(\theta_{(2)})\right] \!=\! A(\theta_{(1)}) \, 
B(\theta_{(2)})\,\d^8(\theta_{(1)}-\theta_{(2)}) , 
\ee
which leads to the same conclusion concerning the degree of divergence
of supergraphs. Despite the presence of non-trivial phase factors,
combinations that involve less than eight chiral derivatives acting on
a $\delta$-function vanish. More details regarding planar supergraph
power counting in the presence of $\star$-products are presented in
appendix \ref {pcr}.

In the case of non-planar supergraphs, the modification to the rules
due to the $\star$-products is more complicated. In particular, the
chiral derivatives can contribute extra factors of momentum to the
loop integrals. Thus, the methods described here only offer a very
poor upper bound on the degree of divergence of non-planar graphs.

\subsection{Analysis of planar $n$-point graphs}

We now explicitly show how to implement the points listed at the
beginning of section \ref {proof}.  As explained in the previous
pages, the superficial degree of divergence of a generic planar
supergraph is zero if all momenta contribute to the loop
integrals. However, since the external legs in a supergraph do not
contribute to these integrals, certain Wick contractions can
potentially give rise to diagrams with positive degree of divergence.
We analyze the contribution of different Wick contractions to generic
supergraphs and, using the Feynman rules of the theory, we explain how
to reduce, in each case, the degree of divergence from that determined
by power counting to a negative value. 

Notice that this analysis is applicable separately to both planar and
non-planar diagrams, but it only allows us to conclude that the planar
supergraphs are finite because for these we have a stronger bound on
the superficial degree of divergence. A limited number of planar
diagrams require special attention and these are discussed in
subsection \ref {special}.

\subsubsection{Graphs involving a cubic vertex}
\label{ccont}

We first examine the case where an external leg is attached to a cubic
vertex. We consider
\be
\la\Phi(z_1)\Phi(z_2)\Phi(z_3) \int\!\dr^{12}z\:i{\cal L}_3(z)\ra\, ,
\ee
where
\be
\label{orderg}
i\,{\cal L}_3=(-g)\,\Tr{\biggl [}\frac{1}{12}\,\fr{\parm}\,
\Phi\,[{\frac{{\bar d}^4}{\parm^2}\Phi},\partial
{\frac{{\bar d}^4}{\parm^2}\Phi}]_\star
+4\,\frac{{\bar d}^4}{\parm^3}\Phi\,[\Phi,{\bar \partial}\Phi]_\star
{\biggr ]}\, .
\ee
We start with the first term in (\ref {orderg}) and show that all the
terms it produces on Wick contraction are finite. Assume leg $1$ is
external while $2$ and $3$ are internal. 

\begin{figure}[!h]
\hsp{1}
\includegraphics[width=7cm]{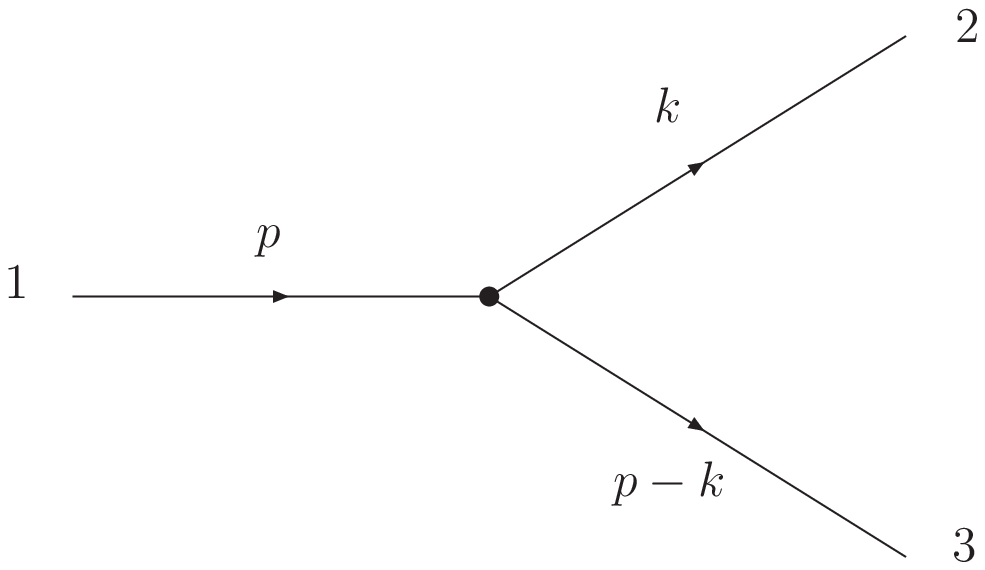}
\label{3ptfig}
\end{figure}

\ndt
We then have the following Wick contractions 
\begin{align}
-\frac{g}{12}\,\Tr^\prime{\biggl \{}&\fr{\parm}\Delta_1[
\frac{{\bar d}^4}{\parm^2}\Delta_2,\partial\,
\frac{{\bar d}^4}{\parm^2}\Delta_3]_\star+(2\leftrightarrow3) \nn \\
&\hsp{-0.47} +\fr{\parm}\Delta_2[\frac{{\bar d}^4}{\parm^2}\Delta_3,
\partial\,\frac{{\bar d}^4}{\parm^2}\Delta_1]_\star
+(2\leftrightarrow3) \nn \\
&\hsp{-0.47} +\fr{\parm}\Delta_2[\frac{{\bar d}^4}{\parm^2}\Delta_1,
\partial\,\frac{{\bar d}^4}{\parm^2}\Delta_3]_\star+\fr{\parm}\Delta_3
[\frac{{\bar d}^4}{\parm^2}\Delta_1,\partial\,
\frac{{\bar d}^4}{\parm^2}\Delta_2]_\star{\biggr \}}\, .
\label{threev1}
\end{align}
The primed trace, which indicates that only the indices associated with
the interaction point are summed over, is further explained in appendix
\ref{wick}. For convenience, we will only explicitly write this trace
in the first of a series of steps.  In the following we will
repeatedly integrate by parts the chiral derivatives that appear in
the contractions above. The rules governing these manipulations are
described in appendix \ref {ssstar}.

In the first term in (\ref {threev1}), the presence of a $\fr{\parm}$
acting on an external leg implies that we lose a factor of momentum
from the denominator of the loop-integral making it potentially
linearly divergent. In the first line, we integrate the ${\bar d}^4$
from leg $2$ for example moving it to leg $1$ (it cannot move to leg
$3$ since ${\bar d}^{\,5}=0$). This takes two powers of momentum out
of the loop-integral, rendering it finite. The second line in
momentum-space is
\bea
\fr{k_-}\fr{{(p_--k_-)}^2}\,\frac{p}{p_-^2}\;\Delta_2
[{\bar d}^4\Delta_3,{\bar d}^4\Delta_1]_\star\, .
\eea
The presence of the factor $p\;\frac{{\bar d}^4}{p_-^2}$ on the
external leg improves the convergence of the integral by a single
power of momentum. Hence this contribution is finite. The third line
in (\ref {threev1}) is more subtle. We start with the first term
and integrate the superspace chiral derivatives from leg $3$ to leg $2$
\bea
\fr{\parm}\Delta_2[\frac{{\bar d}^4}{\parm^2}\Delta_1,\partial\,
\frac{{\bar d}^4}{\parm^2}\Delta_3]_\star
=\frac{{\bar d}^4}{\parm}\Delta_2[\frac{{\bar d}^4}{\parm^2}\Delta_1,
\partial\,\fr{\parm^2}\Delta_3]_\star\, .
\eea
Using the last relation in (\ref {compprop}) we rewrite this as
\bea
-\partial\,\fr{\parm^2}\Delta_3[\frac{{\bar d}^4}{\parm^2}
\Delta_1,\frac{{\bar d}^4}{\parm}\Delta_2]_\star\, .
\eea
The third line in (\ref {threev1}) now reads
\bea
-\partial\,\fr{\parm^2}\Delta_3[\frac{{\bar d}^4}{\parm^2}\Delta_1,
\frac{{\bar d}^4}{\parm}\Delta_2]_\star+\fr{\parm}\Delta_3
[\frac{{\bar d}^4}{\parm^2}\Delta_1,\partial\,
\frac{{\bar d}^4}{\parm^2}\Delta_2]_\star\, .
\eea
Working in momentum space, this becomes
\bea
-\,{\biggl (}\frac{p-k}{{(p_--k_-)}^2}\,\fr{p_-^2}\,
\fr{k_-}\,-\,\fr{p_--k_-}\,\fr{p_-^2}\,\frac{k}{k_-^2}\,{\biggr )}\,
\Delta_3\,[{\bar d}^4\Delta_1,{\bar d}^4\Delta_2]_\star\, .
\eea
When tracking ultra-violet divergences, our focus is on large
loop-momenta. For $k{\gg}p$, the leading term in parentheses
vanishes implying finiteness of this contribution.

Having analyzed diagrams resulting from the first vertex in (\ref
{orderg}) we now move to the second vertex. The Wick contractions in
this case yield
\begin{align}
-4g\,\Tr^\prime{\biggl \{}&\frac{{\bar d}^4}{\parm^3}\Delta_1[\Delta_2,
{\bar \partial}\,\Delta_3]_\star+(2\leftrightarrow3) \nn \\
+&\frac{{\bar d}^4}{\parm^3}\Delta_2[\Delta_3,{\bar \partial}\,
\Delta_1]_\star+(2\leftrightarrow3) \nn \\
+&\frac{{\bar d}^4}{\parm^3}\Delta_3[\Delta_1,{\bar \partial}\,
\Delta_2]_\star+\frac{{\bar d}^4}{\parm^3}\Delta_2[\Delta_1,
{\bar \partial}\,\Delta_3]_\star{\biggr \}}\, .
\end{align}
In line $1$, both internal legs are free of $\bar d$'s. However, they
are both attached to (internal) propagators that carry a factor of
$d^4$ (see equation (\ref {prop})). Integrating this factor of $d^4$
from either internal leg takes it out of the loop integral (since
$d^{\,5}=0$) and ensures convergence. Line $2$ is finite because the
numerator involves factors of ${\bar p}$ (from ${\bar
\partial}\,\Delta_1$), the external momentum, which factors out of the
integral. As in the previous case, line $3$ involves a little work. We
start with the first term
\bea
\frac{{\bar d}^4}{\parm^3}\Delta_3[\Delta_1,
{\bar \partial}\,\Delta_2]_\star=\frac{1}{\parm^3}\Delta_3
[\Delta_1,{\bar \partial}\,{\bar d}^4\,\Delta_2]_\star\, ,
\eea
which is rewritten, using antisymmetry of the $\star$-commutator and
the last relation in (\ref {compprop}), as
\bea
{\bar \partial}\,{\bar d}^4\,\Delta_2[\frac{1}{\parm^3}\Delta_3,
\Delta_1]_\star=-{\bar \partial}\,{\bar d}^4\,\Delta_2[\Delta_1,
\frac{1}{\parm^3}\Delta_3]_\star\, .
\eea
In momentum space, the entire third line is now
\bea
{\biggl (}\frac{{\bar p}-{\bar k}}{k_-^3}-
\frac{\bar k}{{(p_--k_-)}^3}\,{\biggr )}\,{\bar d}^4\Delta_2
[\Delta_1,\Delta_3]_\star\, .
\eea
For $k{\gg}p$ the leading terms cancel and this makes the
resulting integral finite.

\subsubsection{Graphs involving a quartic vertex}
\label{qcont}

We start from
\bea
\la\Phi(z_1)\Phi(z_2)\Phi(z_3)\Phi(z_4)\int\!\dr^{12}z\:
i{\cal L}_4(z)\ra\, ,
\eea
where
\bea
\label{ordergsq}
i\,{\cal L}_4(z)=i\,\frac{g^2}{16}\,\Tr{\biggl
\{}\fr{\parm}[\Phi,\parm\Phi]\fr{\parm}[{\frac{{\bar d}^4}
{\parm^2}\Phi},{\frac{{\bar d}^4}
{\parm}\Phi}]_\star+\fr{2}[\Phi,\frac{{\bar d}^4}
{\parm^2}\Phi]_\star[\Phi,\frac{{\bar d}^4}
{\parm^2}\Phi]_\star{\biggr \}}\, .  
\eea
We will analyze separately the cases in which either one or two legs
of the quartic vertex are external. In the following, we will ignore
the overall factor of $i\frac{g^2}{16}$.

\vskip 0.3cm

\ndt {\bf{Graphs involving two external lines}}

\vskip 0.2cm
\ndt
We choose legs $1$ and $2$ to be external while $3$ and $4$ are
internal. 

\begin{figure}[!h]
\hsp{1}
\includegraphics[width=7cm]{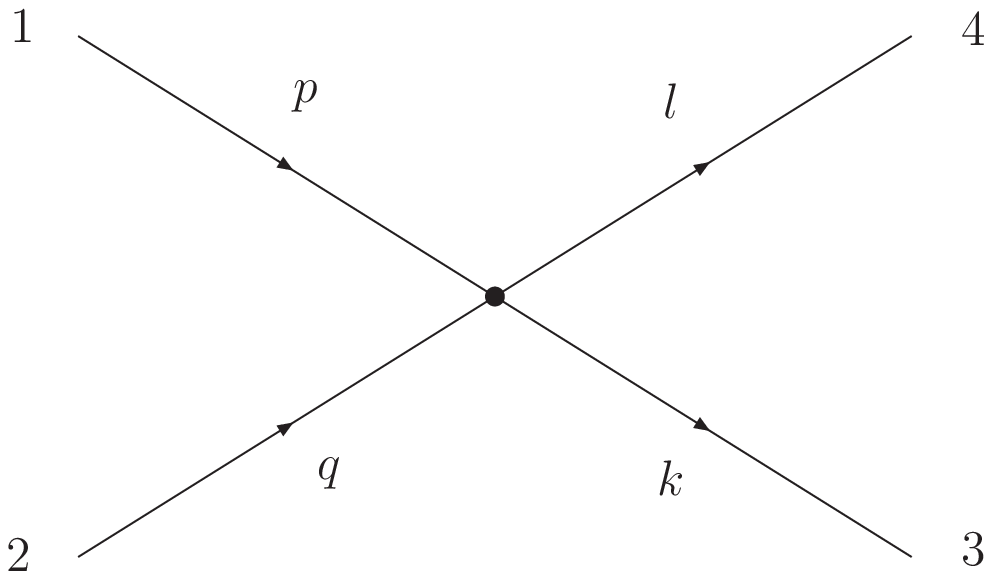}
\label{4pt1fig}
\end{figure}

\ndt
The first quartic vertex in (\ref {ordergsq}) gives rise to twenty-four
contractions. The analysis of the majority of these involves the same
manipulations utilized in the case of the cubic vertex: superspace
chiral derivatives, $d$'s or $\bar d$'s, are integrated by parts from
an internal leg onto an external one to remove factors of momentum
from the numerator of loop integrals. The only noticeable difference
with respect to the analysis in the previous section is that the
integrations by parts can produce non-trivial phase factors which,
however, do not affect the ultra-violet behavior of the diagrams. For
completeness we discuss these contractions in appendix \ref {details},
focussing instead here on those which require special attention. These
are
\be
\hsp{-0.2}\Tr^\prime \left\{\fr{\parm}[\Delta_1,\parm\,\Delta_3]_\star\,
\fr{\parm}[\frac{{\bar d}^4}{\parm^2}\Delta_2,\frac{{\bar d}^4}{\parm}
\Delta_4]_\star +(1\leftrightarrow2)+(3\leftrightarrow4)
+(1\leftrightarrow2,3\leftrightarrow4)\right\}.
\label{fourv1}
\ee
In terms of momenta, the first term becomes
\bea
\label{line6}
\frac{k_-}{k_-+p_-}\,\fr{q_-+l_-}\,\fr{q_-^2}\,\fr{l_-}\,
[\Delta_1,\Delta_3]_\star[{\bar d}^4\Delta_2,{\bar d}^4\Delta_4]_\star\, .
\eea
The potentially divergent contribution in this expression is cancelled
by contributions from the second quartic vertex in (\ref
{ordergsq}). The twenty-four contractions from the second quartic
vertex reduce to twelve terms due to the symmetry in the
expression. This symmetry factor cancels the $\fr{2}$ in front of the
vertex. The contractions that cancel the divergences in (\ref{fourv1})
can be written as
\bea
[\Delta_1,\frac{{\bar d}^4}{\parm^2}\Delta_3]_\star
[\Delta_4,\frac{{\bar d}^4}{\parm^2}\Delta_2]_\star+
(1\leftrightarrow2)+(3\leftrightarrow4)+(1\leftrightarrow2,
3\leftrightarrow4)\, .
\eea
We integrate the ${\bar d}^4$ from leg $3$ to leg $4$ (if a single
$\bar d$ hits the external leg $1$, the integral is rendered
finite). In momentum space, the first term is
\bea
\fr{k_-^2}\,\fr{q_-^2}\,[\Delta_1,\Delta_3]_\star[{\bar d}^4
\Delta_4,{\bar d}^4\Delta_2]_\star\, .
\eea
In the large loop-momentum limit $k,l\,{\gg}\,p,q$ the leading order
term exactly cancels against that in (\ref {line6}) implying that the
combined contribution is finite. In appendix \ref {details} we present
the finiteness analysis for the remaining contractions involving the
second quartic vertex.

\vskip 0.3cm

\ndt {\bf{Graphs involving one external line}}

\vskip 0.2cm
\ndt
We choose leg $1$ to be external keeping $2$, $3$ and $4$ internal. 

\begin{figure}[!h]
\hsp{1}
\includegraphics[width=7cm]{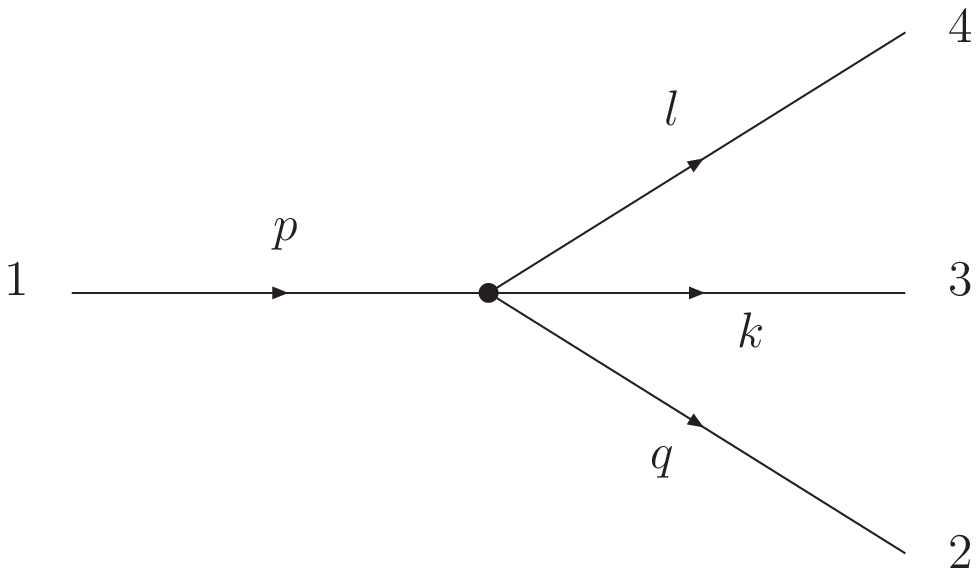}
\label{4pt2fig}
\end{figure}

\ndt
We split our analysis into two portions. We consider first graphs in
which the external leg $1$ does not have a factor of ${\bar d}^4$ on
it. In this case, the contractions for the first quartic vertex in
(\ref {ordergsq}) are
\begin{align}
&\Tr^\prime \left\{\fr{\parm}[\Delta_1,\parm\,\Delta_2]_\star\,
\fr{\parm}[\frac{\bar d^4}{\parm^2}\Delta_3,\frac{\bar d^4}{\parm}
\Delta_4]_\star+\fr{\parm}[\Delta_2,\parm\,\Delta_1]_\star\,
\fr{\parm}[\frac{\bar d^4}{\parm^2}\Delta_3,\frac{\bar d^4}{\parm}
\Delta_4]_\star \right. \nn \\
&\hsp{0.38}+\fr{\parm}[\Delta_1,\parm\,\Delta_2]_\star\,\fr{\parm}
[\frac{\bar d^4}{\parm^2}\Delta_4,\frac{\bar d^4}{\parm}\Delta_3]_\star+
\fr{\parm}[\Delta_2,\parm\,\Delta_1]_\star\,\fr{\parm}
[\frac{\bar d^4}{\parm^2}\Delta_4,\frac{\bar d^4}{\parm}\Delta_3]_\star \nn \\
&\left.\hsp{0.37} +({\rm permutations ~ of~} 2,3,4)\rule{0pt}{14pt} 
\right\} \, . \label{fourv1b}
\end{align}
The contractions from the second quartic vertex in (\ref {ordergsq})
read
\begin{align}
&[\Delta_1,\frac{1}{\parm^2}\Delta_2]_\star[{\bar d}^4\Delta_3,
\frac{{\bar d}^4}{\parm^2}\Delta_4]_\star+[\Delta_1,
\frac{1}{\parm^2}\Delta_2]_\star[{\bar d}^4\Delta_4,
\frac{{\bar d}^4}{\parm^2}\Delta_3]_\star \nn \\
&+({\rm permutations ~ of~} 2,3,4)\, , \label{fourv2b}
\end{align}
where we have integrated the ${\bar d}^4$ away from leg $2$ to legs
$3$ and $4$ (if they move to leg $1$, the integral is finite). In
momentum space, the sum of equations (\ref {fourv1b}) and (\ref
{fourv2b}) is proportional to
\be
\frac{(p_-+k_-){(l_-+q_-)}^2(l_--q_-)+q_-k_-^3-q_-k_-^2p_-
-k_-^3l_-+l_-k_-^2p_-}{(p_-+k_-)(l_-+q_-)l_-^2q_-^2k_-^2}\, .
\ee
Using momentum conservation, in the large loop-momentum limit, the
dominant terms in this expression behave as 
\be
\frac{(k_-+2l_-)p_-^2}{k^3_-l_-^2(k_-+l_-)^2} \, ,
\label{leadterm}
\ee
implying that the graph is finite.

The final case is when the external leg has a factor of ${\bar d}^4$
on it. When studying a complicated supergraph, if we locate a single
external leg (free of ${\bar d}^4$) attached to a three or four-point
vertex, finiteness follows based on the arguments presented so
far. Thus the only cause for concern is a supergraph which has factors
of ${\bar d}^4$ on all its external legs. If this is the case, then in
order for the expression to survive the integration over the entire
measure $\int\,d^4{\theta}d^4{\bar \theta}$ we necessarily need a
factor of $d^4$ to compensate the ${\bar d}^4$. This factor of $d^4$
must come from the internal structure of the supergraph and ensures
that the integrals are rendered finite.

\subsubsection{Exceptional cases}
\label{special}

\begin{figure}[!htb]
\begin{center}
\includegraphics[width=6cm]{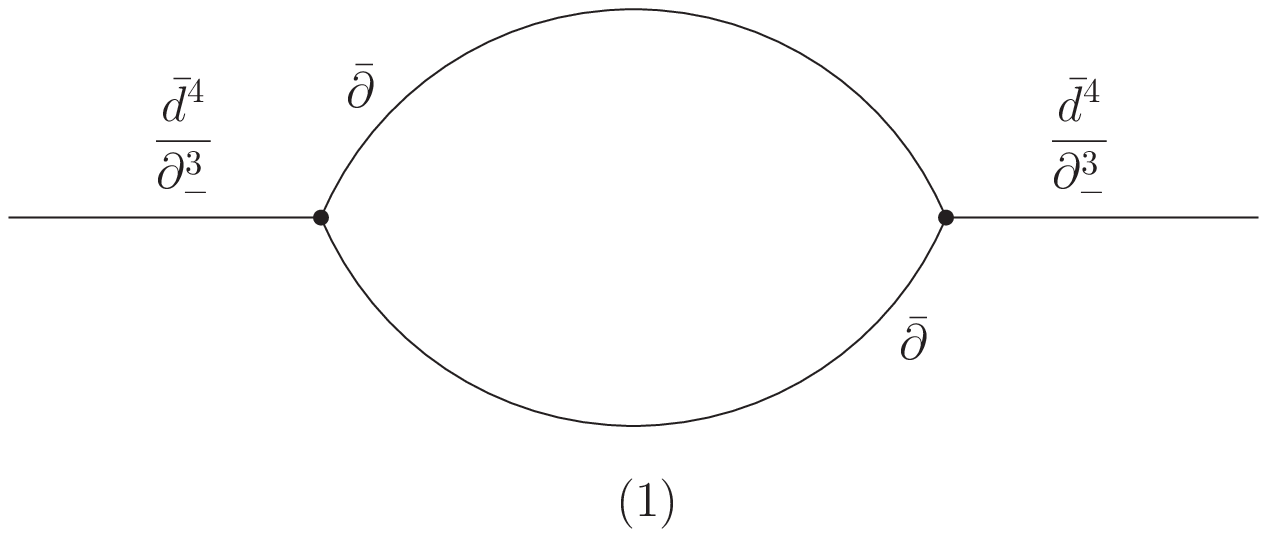} \hsp{1}
\includegraphics[width=6cm]{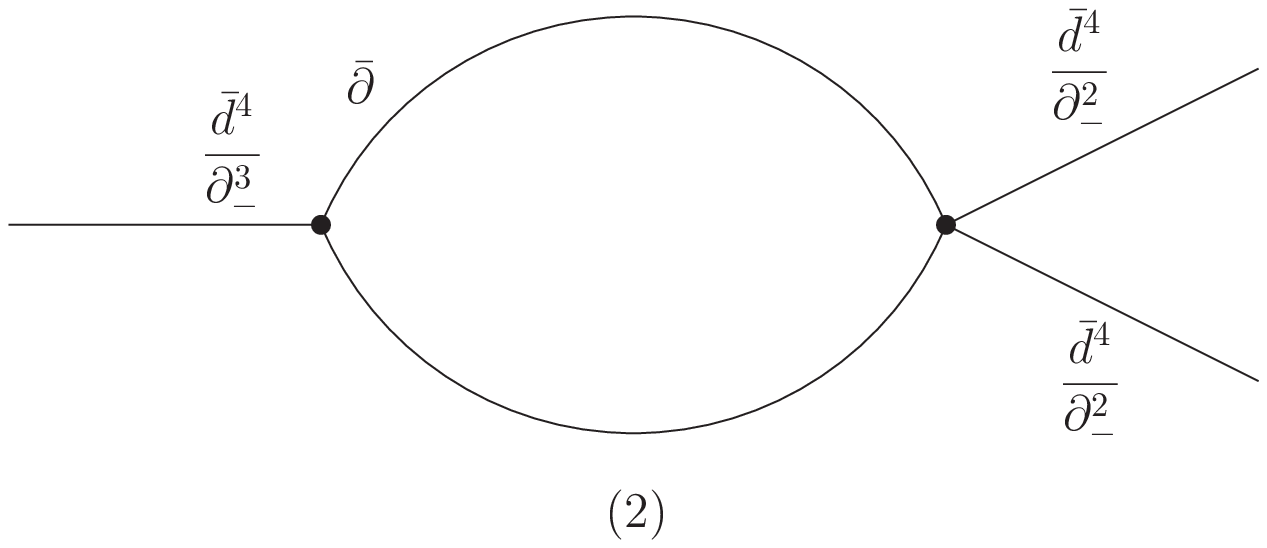} \rule{0pt}{3cm} \\ 
\includegraphics[width=6cm]{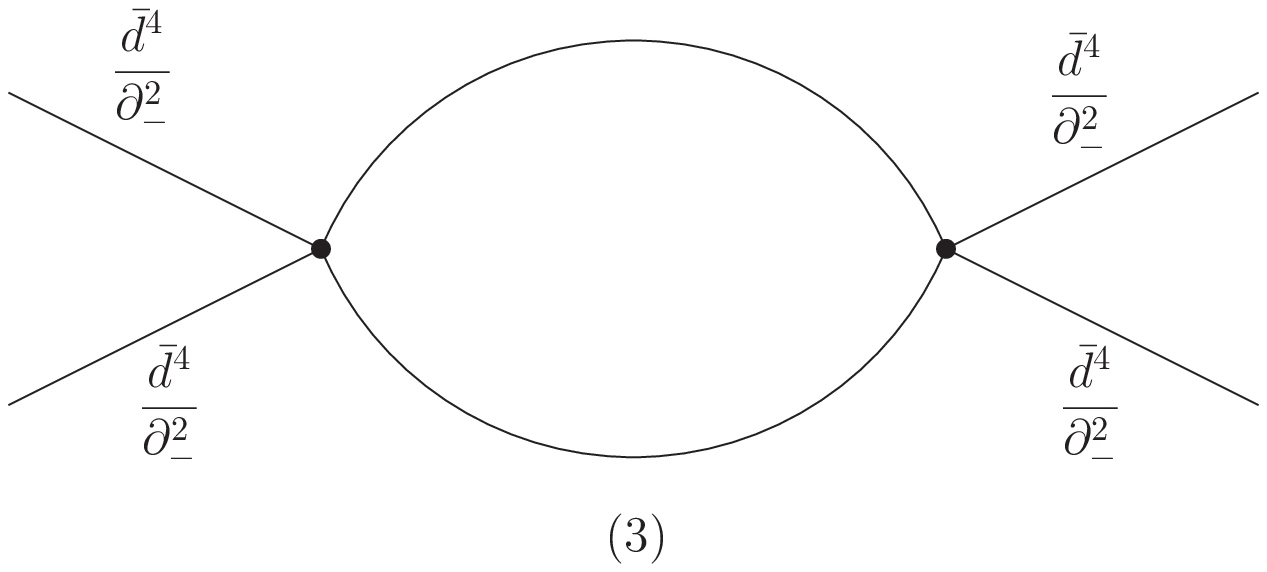} \hsp{1}
\includegraphics[width=6cm]{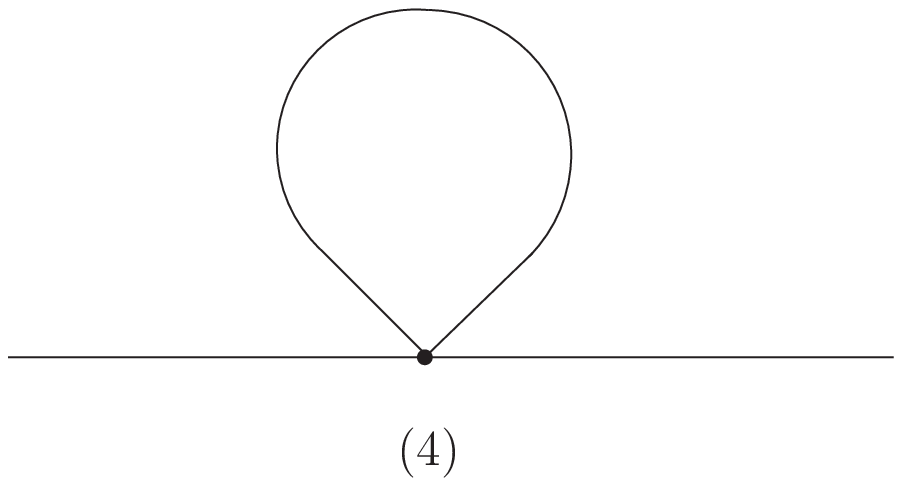} \rule{0pt}{3.5cm}
\vsp{-0.4}
\end{center}
\caption{Diagrams to be treated separately.}
\label{exfig}
\end{figure}

There are a few planar supergraphs that require special attention.
These are depicted in figure \ref {exfig}.
 
\ndt 
The analysis of section \ref {proof} is insufficient to prove
finiteness for some specific contractions corresponding to diagrams of
the types (1), (2) and (3) in figure \ref{exfig}. The chiral
derivative structure in these contractions is explicitly shown in the
figures. Notice that these graphs have factors of $\bar d^4$ acting on
all the external legs and thus they belong to the class discussed at
the end of the previous subsection. Our proof assumes that factors of
$d^4$ can always be integrated by parts from internal to external
lines for diagrams of the type in figure \ref{exfig}. In the present
cases, however,  after the first $d^4$ has been moved out of the loop,
the second factor of $d^4$ can also move to the other internal line
which now has no $d$'s acting on it. The degree of divergence at this
stage is still logarithmic. The resolution to this comes from the
chiral derivative structure. The $\theta$-integrals in all these
graphs trivially vanish in the planar limit since there are not enough
chiral derivatives acting on the $\delta$-functions.

The diagram (4) in figure \ref{exfig} involves a self-contraction
which is also not dealt with in our arguments of the previous
section. In addition, there are subtleties associated with defining
this graph in superspace in the presence of $\star$-products.
Resorting to a simple component calculation, however, one can verify
that in the planar limit the one-loop two-point function of the
$\b$-deformed theory is identical to that in $\calN=4$ Yang--Mills.
Thus the one-loop two-point function which includes the contribution
(4) is finite and has the correct asymptotic behavior at large
momentum.

\begin{center}
* ~ * ~ *
\end{center}

\ndt
We have thus shown that all supergraphs have a negative superficial
degree of divergence. This analysis applies equally well to all
subgraphs within a given supergraph. This allows us to use Weinberg's
theorem~\cite{SW} to conclude that all the Green functions of the
theory are finite. Notice that Weinberg's theorem in its original form
requires Euclidean signature and therefore there are potential
subtleties when using it in light-cone gauge. In our formalism, the
Wick rotation into Euclidean space is permitted thanks to the residual
gauge freedom~\cite{BPP}, which allows us to choose the pole structure
${(p_-+i\epsilon p_+)}^{-1}$ for the operator $\fr{\parm}$~\cite{SM}.
We also point out that Weinberg's theorem has been generalized to
Lorentzian signature in \cite{Z}.

\section{Conclusions}
\label{concl}

In this paper we have studied the finiteness properties of a special
example of $\b$-deformed $\calN=4$ Yang--Mills theory, involving a
single real deformation parameter. Theories in this class, despite the
reduced amount of supersymmetry, preserve many of the remarkable
properties of the parent $\calN=4$ theory.  Our methods show that this
particular $\b$-deformed $\calN=4$ SYM is conformally invariant in the
planar limit. The essential ingredient of this analysis was the
realization that the deformed theory, despite having only $\calN=1$
supersymmetry, could still be formulated in $\calN=4$ light-cone
superspace using suitably defined superspace $\star$-products.  In
this formulation the deformation preserves the ultra-violet 
behavior of the $\calN=4$ theory, in the planar limit, thanks to the
properties of the superspace $\star$-product, which allowed us to
prove the finiteness of all the Green functions in the theory
following the same steps previously utilized in~\cite{SM,BLN2} in the
$\calN=4$ case. 

The results in this paper are valid to all orders in planar
perturbation theory. Instanton effects, which have been studied in the
$\b$-deformed theory in~\cite{vvk}, generalizing previous work done in
$\calN=4$ SYM~\cite{instN4}, are exponentially suppressed in the
planar approximation and therefore cannot spoil the conformal
invariance of the theory in this limit. However, one of the remarkable
features of the $\b$-deformed theory is that it inherits from
$\calN=4$ SYM a modified form of S-duality~\cite{dhk}. Instantons are
expected to play a crucial role in the realization of this symmetry.

The proof presented here is valid for arbitrary choice of the gauge
group, but the case of SU($N$) is of special interest in the context
of the AdS/CFT correspondence. The $\b$-deformation studied in this
paper is believed to be dual to the supergravity background
constructed in~\cite{LM}. Our arguments, showing that the deformed SYM
theory is conformally invariant to all orders in perturbation theory
in the planar limit, suggest that the SO(4,2) isometries of the
supergravity solution of~\cite{LM} should not be affected by string
tree-level corrections.

Our analysis opens up many venues for generalizations. A
straightforward extension involves further deforming the theory with
the addition of mass terms for the fields in the $\calN=1$ chiral
multiplets. These mass deformations can preserve $\calN=1$
supersymmetry or break it completely, but do not spoil the
ultra-violet finiteness of the theory, although they obviously break
conformal invariance.

As already mentioned the theory considered in this paper belongs to
the class of deformations of $\calN=4$ SYM characterized by the
superpotential (\ref{genbdef}). It has been argued~\cite{LS} that the
generic theory in this family can be rendered finite by imposing a
single relation among the parameters,
\be
\g(g,h,\b) = 0 \, .
\label{cond}
\ee 
Our results show that in the planar approximation the theory involving
only a real deformation parameter, $\b$, is finite without any
conditions on $\b$. This is in agreement with the explicit results
of~\cite{fg,rss,milan}. Moreover it implies that the general condition
for finiteness (\ref{cond}), specialized to the case of a single real
$\b$ parameter and $h=1$, should be identically satisfied in the large
$N$ limit at all orders in the Yang--Mills coupling. We note that a
different argument for the all-order finiteness of this theory in the
planar limit was provided in~\cite{milan}.

Although we have only
discussed the case of real $\b$, the light-cone superspace formalism
is well suited to study the case where the deformation parameter is
made complex. In this case, even in the planar limit, the condition
(\ref{cond}) may remain non-trivial. Therefore our approach should
provide interesting insights into the surface of finite theories
defined by (\ref{cond}).

We also believe that the analysis that we presented can be generalized
to the case of a $\b$-deformation that breaks all the supersymmetries
in the theory~\cite{SF}. This would be extremely interesting because
we would then have a proof of conformal invariance, in the planar
limit, for a non-supersymmetric field theory. These issues are
currently under investigation.

Having shown that the $\b$-deformed theory preserves conformal
invariance, it is natural to study the spectrum of scaling dimensions
of gauge-invariant operators in this model. Various results have been
presented in~\cite{fg,rss,milan,np}, confirming and extending the earlier
analysis of~\cite{bl,bjl}. The problem of computing the spectrum of
scaling dimensions can be efficiently recast as an eigenvalue problem
for the dilatation operator of the theory. In the planar limit on
which we have focussed, in the case of the $\calN=4$ Yang--Mills
theory, the dilatation operator can be related to the Hamiltonian of
an integrable spin chain allowing for the use of powerful techniques
such as the Bethe Ansatz to compute anomalous dimensions~\cite{MZ}. It
may be interesting to generalize some of these techniques for use in
studying the $\b$-deformed theory.

\vskip 0.8cm

\ndt
{\bf Acknowledgments} 

\vsp{0.2}
\ndt
We are grateful to Lars Brink, Pierre Ramond, Matthias Staudacher and
Stefan Theisen for discussions and to Stanley Mandelstam for answering
an email query. We thank Sergey Frolov, Sung-Soo Kim, Takeshi Morita,
Hendryk Pfeiffer and Tadashi Takayanagi for helpful comments. The work
of SK was supported in part by a Marie Curie Intra-European Fellowship
and by the EU-RTN network {\it Constituents, Fundamental Forces and
Symmetries of the Universe} (MRTN-CT-2004-005104).

\vskip 1cm

\appendix 

\section{Properties of Star Products}
\label{starprop}

We present here relevant relations satisfied by the star products.

\subsection{Properties of the component $*$-product}

The $*$-product satisfies the following properties
\bea
\label{compprop}
&&\hsp{-2} A*(B*C) = (A*B)*C \qquad ({\rm associativity})\ \nn \\
&&\hsp{-2} [A,[B,C]_*]_* + [B,[C,A]_*]_* + [C,[A,B]_*]_* = 0 \qquad 
({\rm Jacobi ~ identity}) \nn \\
&&\hsp{-2} A*B = AB \qquad ({\rm for} ~ Q_A \cdot Q_B=0 ~ {\rm or} ~
Q_A+Q_B=0) \nn \\
&&\hsp{-2} \Tr(A[B,C]_*) = \Tr(B[C,A]_*)=\Tr(C[A,B]_*)
\qquad ({\rm for} ~ Q_A+Q_B+Q_C=0)\, .
\eea

\subsection{Properties of the superspace $\star$-product}
\label{ssstar}

The superspace $\star$ is essentially a superspace realization of the
component $*$. Naturally it also satisfies all the properties listed
above. The presence of $\star$-products in superspace expressions
modifies the rules for partial integration of chiral derivatives. In
this appendix, we describe manipulations of chiral derivatives. In
the following, the index on the chiral derivatives refers to their
flavor~\footnote{This notation should not be confused with that in the
main text where $d^4$ was used to denote the product of all four
chiral derivatives.}.

In general, the integration by parts of chiral derivatives in
superspace expressions involving $\star$-products gives rise to phase
factors. This is a consequence of the modification of the standard Leibniz
rule, which, in the presence of $\star$-products, becomes
\be
\label{leib}
d^1(F\star G)=\:d^1F\star
(\er^{-i\pi\b\,Q^1}G)+(\er^{i\pi\b\,Q^1}F)\star d^1G\, .
\ee
Similar relations hold for the other chiral derivatives. In this
relation, $Q^1$ is an operator, which  acts differently on the various
terms in the expansion of the superfields $F$ and $G$. Therefore, in
order to prove (\ref{leib}), it is convenient to decompose the
superfields into pieces which have definite flavor charge. This is
achieved by the standard $\theta$-expansion
\bea
&&F=f_{(0,0)}+f_{(1,0)m}\theta^m+f_{(0,1)}^m{\bar \theta}_m+
\cdots \nn \\
&&G=g_{(0,0)}+g_{(1,0)m}\theta^m+g_{(0,1)}^m{\bar \theta}_m+
\cdots \ . 
\label{thexp}
\eea
We then have
\bea
d^1\,(f \star g)=d^1\,[f\,g\,e^{i\pi\b(Q^1_fQ^2_g-Q^2_fQ^1_g)}]\, ,
\label{leicomp}
\eea
where the terms in the exponential are now numbers and $f$ and $g$ are
generic terms in the expansions (\ref{thexp}). In (\ref{leicomp}) we 
can now use the ordinary Leibniz rule to obtain
\bea
\label{a6}
d^1\,(f \star g)=(d^1\,f)\,g\,e^{i\pi\b(Q^1_fQ^2_g-Q^2_fQ^1_g)}
+f\,(d^1\,g)\,e^{i\pi\b(Q^1_fQ^2_g-Q^2_fQ^1_g)}\, .
\eea
From the definition of the supercharges in (\ref {charges}) and table
\ref{thetacharges} it follows that
\bea
Q^1_{d^1f}\,=\,Q^1_f \quad {\rm and} \quad Q^2_{d^1f}\,=\,Q^2_f+1\, ,
\eea
implying that (\ref {a6}) can be rewritten as
\bea
d^1\,(f \star g)=(d^1\,f)\,\star\,g\,e^{-i\pi\b\,{Q^1_g}}
+f\,\star\,(d^1g)\,e^{i\pi\b\,{Q^1_f}}\, .
\eea
We now re-sum the component pieces into superfields and this yields
(\ref {leib}). 

In manipulating superspace expressions it is often necessary to
integrate by parts multiple chiral derivatives. This is easily achieved
by repeatedly using (\ref{leib}). In particular, for
\be
d^1d^2d^3d^4\,(F\star G)\, ,
\ee
we obtain
\bea
\label{newprop}
d^1d^2d^3d^4\,(F\star G)&\!\!=\!\!&(d^1d^2d^3d^4\,F)\star G-\er^{p_1}\,
(d^2d^3d^4\,F)\star d^1G+\er^{p_2}\,(d^3d^4\,F)\star d^2d^1\,G \nn \\
&&+\cdots+F\star d^1d^2d^3d^4G\, .
\eea
The exact values of the phase factors, $\er^{p_1}$, $\er^{p_2}$,
$\ldots$, which can be computed using (\ref {leib}) are not essential
to our analysis. This is because our proof, in section \ref {proof},
relies on moving four chiral derivatives which does not produce a
phase. That is, partially integrating four chiral derivatives, ensures
that the complicated phase factors cancel each other. This is because
the product of all four chiral derivatives is uncharged under the
U(1)$\times$U(1) flavor symmetry.  In particular, for three generic
superfields $F$, $G$ and $H$, the Leibniz rule implies that
\bea
\label{keyprop}
\int\dr^{12}z\,\{(d^1d^2d^3d^4\,F)\star G\}\,d^1d^2d^3d^4\,H 
= \int\dr^{12}z\,\{F\star (d^1d^2d^3d^4 G)\}\,d^1d^2d^3d^4\,H\, ,
\eea
since the chiral derivatives cannot move from $F$ to $H$. Equation
(\ref {keyprop}) is non-trivial to prove directly in superspace
because the superfields do not carry definite U(1)$\times$U(1)
charge. However, the result is straightforward to verify, by
decomposing the (generic) superfields into pieces which carry definite
flavor charge, as was done in the derivation of the modified Leibniz
rule (\ref{leib}).

\section{Wick Contraction Notation}  
\label{wick}

The superfield propagator is  
\bea
\la{(\Phi)}^u_v(z_1)\,{(\Phi)}^r_s(z_2)\ra=\la\Phi^a\,{(T^a)}^u_v
\Phi^b\,{(T^b)}^r_s\ra=\Delta^{ur}_{vs}\,(z_1-z_2)\, , 
\eea 
A sample Wick contraction is 
\be
\bigg(\Phi^{u_1}_{v_1}(z_1)\,\Phi^{u_2}_{v_2}(z_2)\,
\Phi^{u_3}_{v_3}(z_3)\,\Phi^{u_4}_{v_4}(z_4)\bigg)
\raisebox{-13pt}{\hspace*{-1.55cm}
\rule{0.4pt}{5pt}\rule{1.85cm}{0.4pt}\rule{0.4pt}{5pt}}
\hspace*{-0.5cm} 
\bigg([\Phi(z),
\raisebox{-17pt}{\hspace*{-4.16cm}
\rule{0.4pt}{9pt}\rule{4.71cm}{0.4pt}\rule{0.4pt}{9pt}}
\hspace*{-0.71cm} 
\frac{{\bar d}^4}{\parm^2}\Phi(z)]_\star\bigg)^r_{\!\!s}
\!\bigg(
\raisebox{-21pt}{\hspace*{-7.75cm}
\rule{0.4pt}{13pt}\rule{7.83cm}{0.4pt}\rule{0.4pt}{13pt}}
\hspace*{-0.23cm}
[\Phi(z),
\raisebox{-25pt}{\hspace*{-10.13cm}
\rule{0.4pt}{17pt}\rule{10.7cm}{0.4pt}\rule{0.4pt}{17pt}}
\hspace*{-0.67cm}
\frac{{\bar d}^4}{\parm^2}\Phi(z)]_\star
\bigg)^s_{\!\!r}\, , 
\ee
where 
\bea
([\Phi(z),\frac{{\bar d}^4}{\parm^2}\Phi(z)]_\star)^r_s=
{\Phi}^r_m(z)\star\frac{{\bar d}^4}{\parm^2}{\Phi}^m_s(z)- 
\frac{{\bar d}^4}{\parm^2}{\Phi}^r_m(z)\star{\Phi}^m_s(z)\, .  
\eea 
We treat the propagator between superspace points $z$ and $z_1$ as a
matrix in the indices associated with the point $z$. We simplify our
notation considerably by not explicitly showing the $z_1$ indices and
the dependence on $(z-z_1)$. We write
\bea
\Delta^{s\;u_1}_{r\;v_1}(z-z_1)\equiv(\Delta_1)_r^s\equiv\Delta_1\, .
\eea
In this new notation, the contraction reads
\be
\Tr^\prime\left\{[\Delta_1,\Delta_2]_\star\;[\Delta_3,
\Delta_4]_\star\right\} \, ,
\ee 
where the symbol $\Tr^\prime$ refers to the fact that only the indices
associated with the point $z$ are contracted.

\section{Planar supergraphs and power counting}
\label{pcr}

\begin{figure}[!htb]
\begin{center}
\includegraphics[width=4cm]{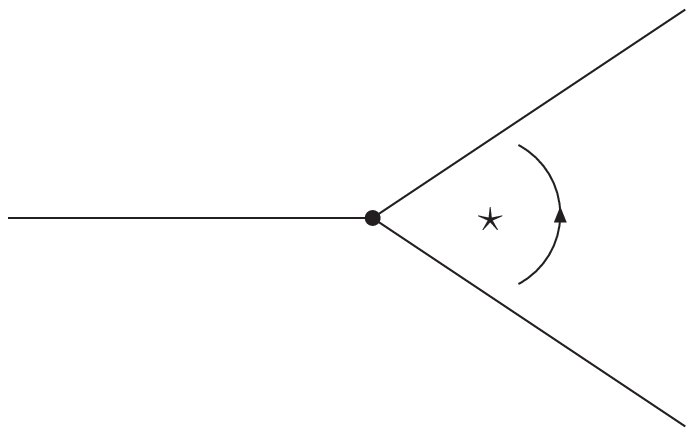} \hsp{2} 
\includegraphics[width=4cm]{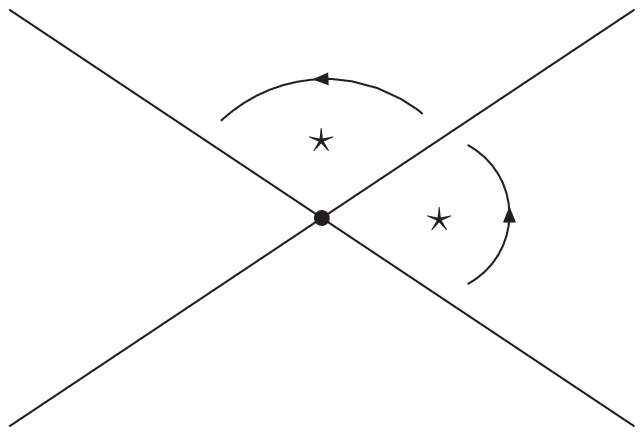}
\end{center}
\vsp{-0.4}
\caption{Star-products in supergraphs.}
\label{starvert}
\end{figure}

This appendix illustrates some basic manipulations of planar
supergraphs in the presence of $\star$-products. The $\star$-products
acting at three- and four-point vertices are shown explicitly in
figure \ref{starvert}. The arrows on the $\star$'s between two lines
refer to the order in which the corresponding superfields are
multiplied.

\ndt
Planar supergraphs in the $\b$-deformed theory are characterized by
the fact that all these $\star$-products that act between adjacent
legs in a vertex have the same orientation. In other words the
$\star$'s all act either clockwise or counter-clockwise. Using the
properties listed in appendix \ref {starprop} these $\star$'s can be
moved around the vertex as long as their orientation is
preserved. This property is used in the procedure of step-wise
integration over the fermionic variables as explained in section 
\ref{sgpc}.

\begin{figure}[!htb]
\begin{center}
\includegraphics[width=4.7cm]{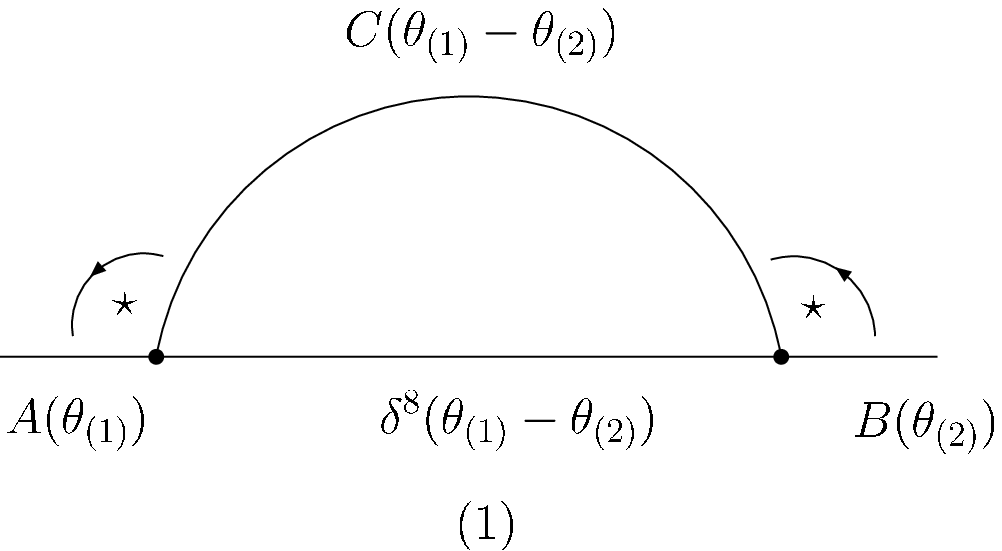} \hsp{0.15}
\includegraphics[width=4.7cm]{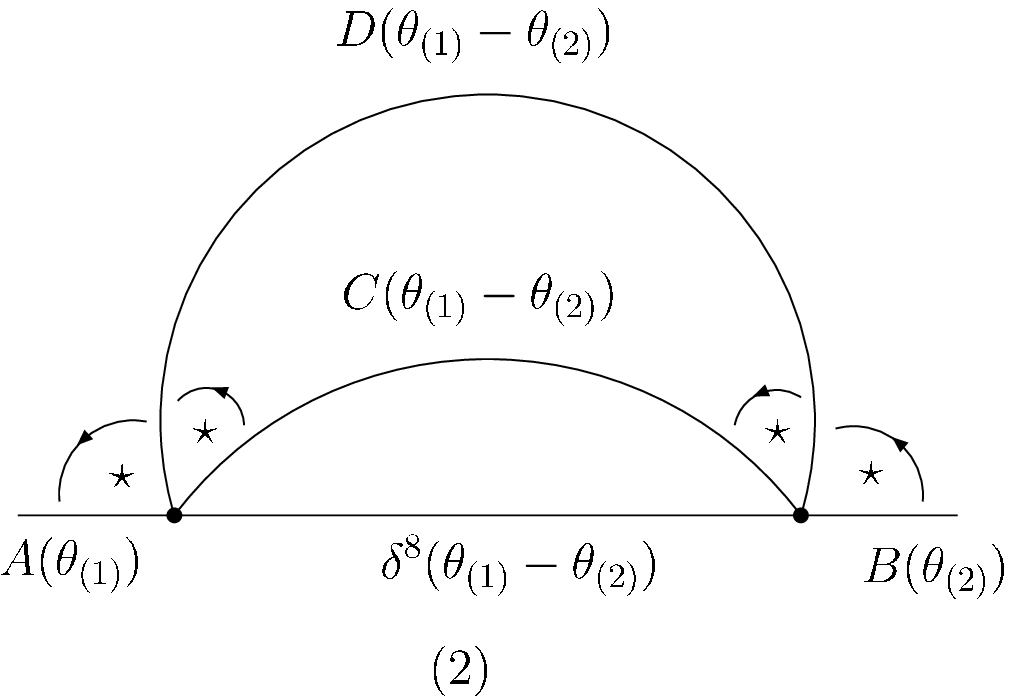} \hsp{0.15}
\includegraphics[width=4.7cm]{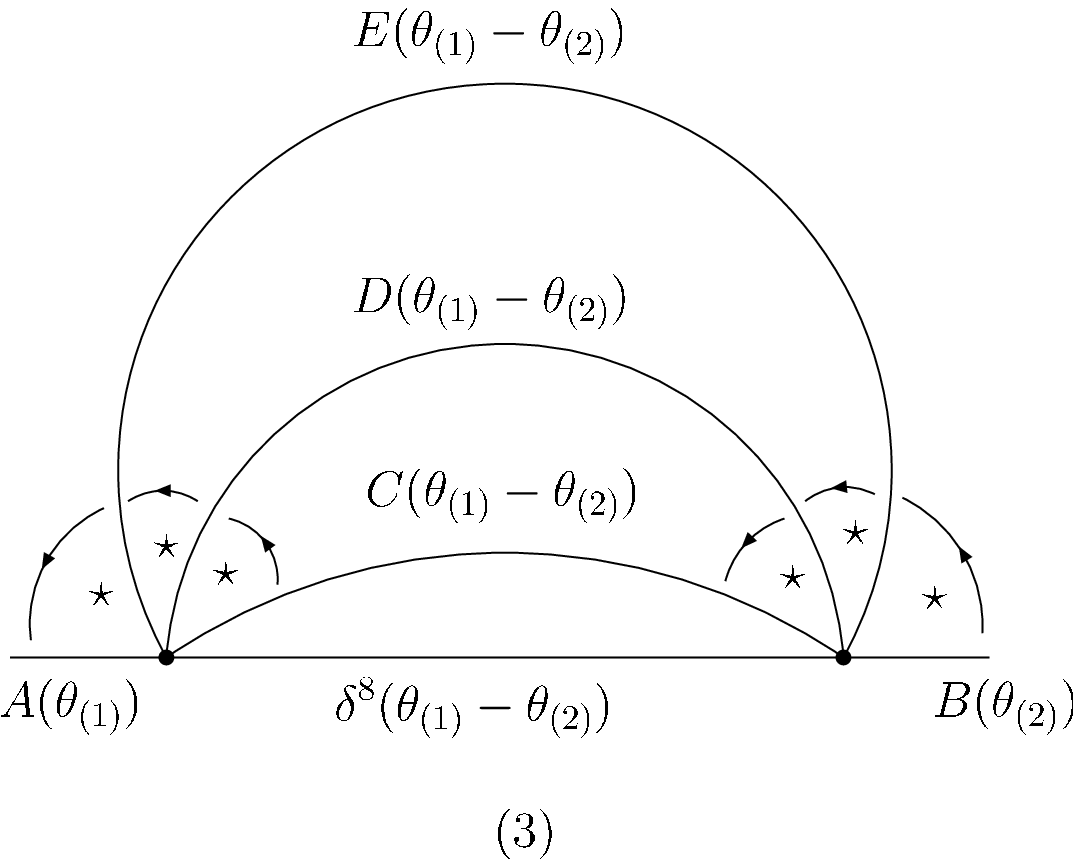}
\end{center}
\vsp{-0.5}
\caption{Loops in $\theta$-space produced by the step-wise fermionic 
integrations.}
\label{figonion}
\end{figure}

The procedure of step-wise integration over the $\theta$'s using the
$\d$-functions in the superfield propagators shrinks internal lines in
a supergraph and can result in self-contracting vertices. These pose a
potential problem since their correct definition in the presence of
$\star$-products is rather subtle.  However, this type of vertex can
be avoided by carefully choosing the order in which the
$\theta$-integrals are performed. It is easy to verify that
self-contractions only arise when shrinking the internal lines in
graphs of the type shown in figure \ref{figonion}.  Therefore we
explain below how to treat generic graphs in this class. Notice that
self-contracting lines can also be induced by the Feynman rules
before the shrinking process is initiated. These primitive
self-contracting vertices are treated as explained in subsection \ref
{special}.

\ndt 
In all these diagrams, the approach is the same. We illustrate the
method in the case of the diagram (2) in the figure (note that the
diagram (1) is simply dealt with by using formula (\ref {grsN42})).
Using the fact that the $\star$-product is associative, we organize
the order of $\star$-products of fields so that at point 1 we have
$C{\star}(D{\star}A)$ and at point 2 $(B{\star}D){\star}C$. We now
$\theta$-expand legs $A$, $B$ and $D$ into terms of definite charge
under U(1)$\times$U(1). The charges carried by the incoming and
outgoing legs $A$ and $B$ are equal and opposite. Applying charge
conservation along a given internal line, for example $C$, we see that
for each SU(4) flavor, the variables $\theta_{(1)}$ and $\theta_{(2)}$
pick up equal phases. It is clear that such a procedure applies to
more complicated cases such as (3) in figure \ref{figonion}. 

When dealing with graphs that have more external legs, it is useful to
view these external legs as a single block when applying the analysis
described above. It is important to note that this ``block" usually
contains $\star$'s in it and hence has a non-trivial dependence on
$\beta$.

In some cases, manipulations on supergraphs can lead to an uneven
(non-singlet) combination of the chiral derivatives. In these
situations it might appear that charge conservation previously used
may be violated. However these uneven distributions of the chiral
derivatives always appear from partial integrations and hence the
resulting vertices are always accompanied by phase factors as
explained in appendix \ref{ssstar}. These phase factors cancel against
those from the apparent violation of charge conservation as a
consequence of the modified Leibniz rule (\ref{leib}). This implies
that in similar situations the $\star$-products can be evaluated
assuming that the  $\star$ does not act on the $d$'s. This simple rule
is valid for any effective vertex arising from our procedure. The
methods applied to graphs having an even distribution of chiral
derivatives therefore are also applicable here.

\section{Planar versus non-planar supergraphs}
\label{pvnp}

In this appendix, we explicitly illustrate the difference between
planar and non-planar graphs in the context of power counting. This
difference, best illustrated with the two-point function, is explained
in the case of a specific Wick contraction.

Our starting point is the non-planar graph shown in figure 4. We will
explain why the power counting rules described in subsection
\ref{sgpc} and in appendix \ref{pcr} are less useful in this
case. Having done this, we turn to the planar case and show why the
same power counting rules work in that case exactly as with $\calN=4$
Yang--Mills.

\begin{figure}[!htb]
\begin{center}
\includegraphics[width=6cm]{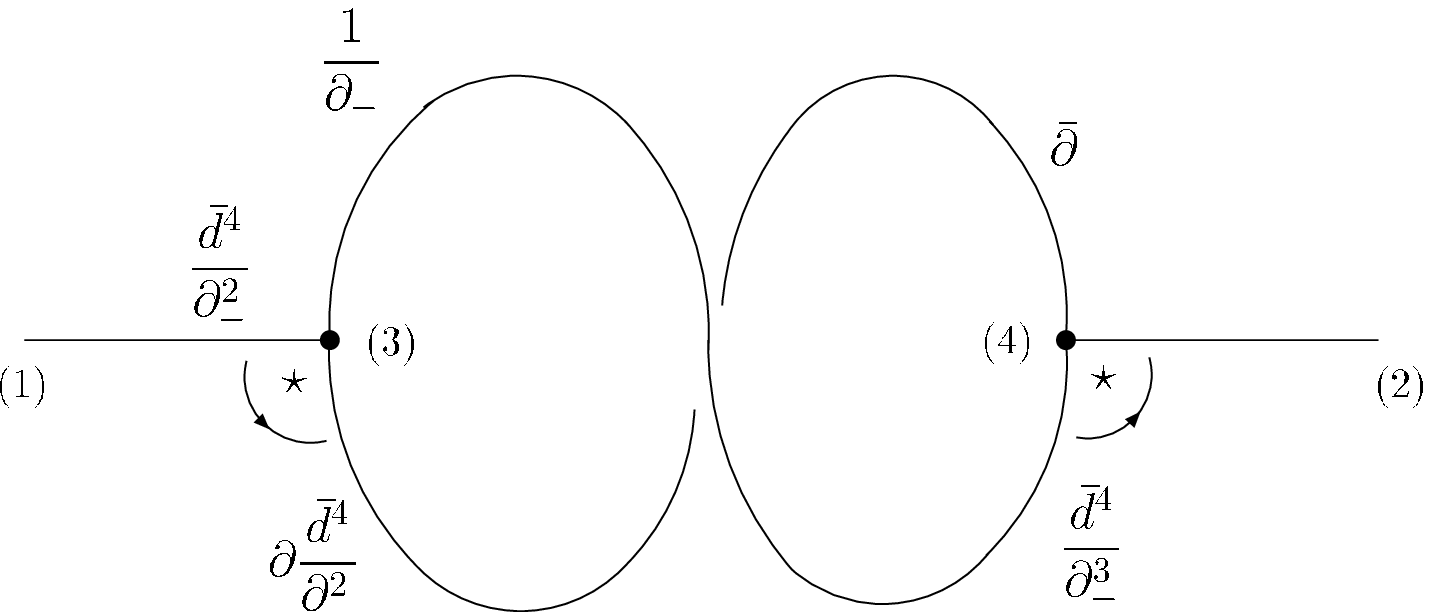} 
\hsp{0.5} \raisebox{1.07cm}{$\Rightarrow$} \hsp{0.5}
\includegraphics[width=6cm]{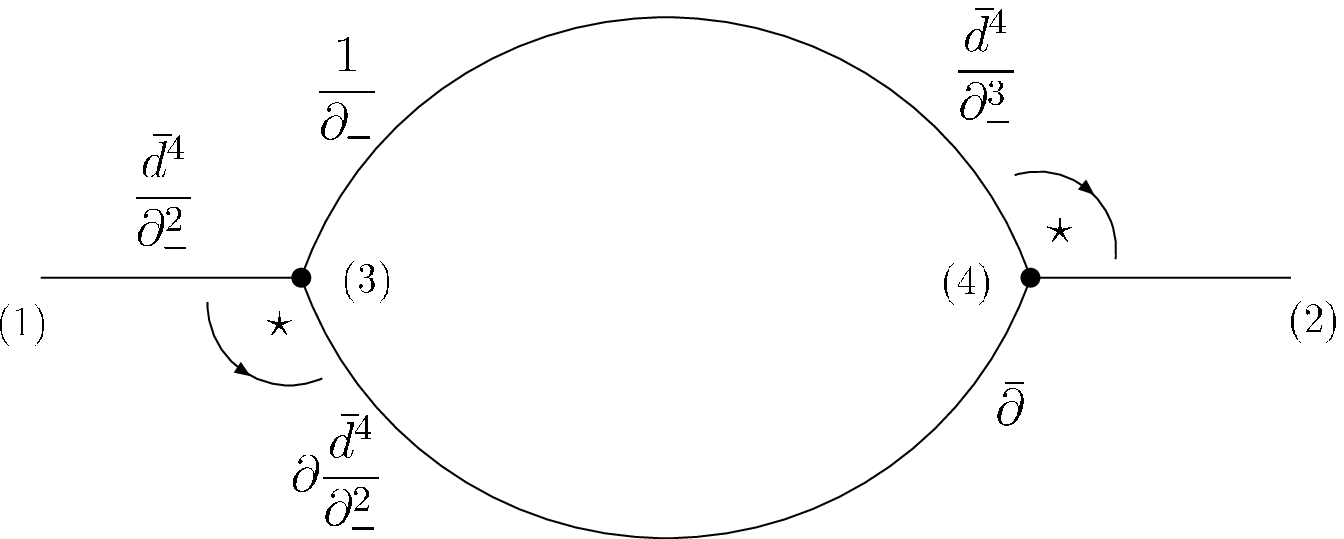} 
\end{center}
\vsp{-0.4}
\caption{Non-planar contribution to the one-loop two-point function.}
\label{2pt-pl}
\end{figure}

\ndt
Reading off Feynman rules from the action (\ref {ans}), we see that
figure 4 is proportional to~\footnote{Some of the equations in this
appendix contain single chiral derivatives. To avoid ambiguities in
the notation we therefore denote the product of four chiral or
anti-chiral derivatives respectively by $[d_{(i)}]^4$ and  $[\bar
d_{(i)}]^4$, where the subscript $(i)$ refers to the superspace point.}
\bea
\label{nptp}
&&\int \dr^4\theta_{(3)}\,\dr^4\bar\theta_{(3)}\,\dr^4\theta_{(4)}
\,\dr^4\bar\theta_{(4)}\,\dr^4k\:\frac{(p-k)(\bar p-\bar k)}
{p_\mu^4p_-^2k_\nu^2{(p_--k_-)}^2{(p_\rho-k_\rho)}^2}\;
\delta^8(\theta_{(3)}-\theta_{(4)})  \\
&&\times \,[d_{(1)}]^4[\bar d_{(1)}]^4\delta^8(\theta_{(1)}-
\theta_{(3)})\,\star_3\,[\bar d_{(3)}]^4[d_{(3)}]^4
\delta^8(\theta_{(3)}-\theta_{(4)})\,\star_4^{-1}\,[d_{(4)}]^4
\delta^8(\theta_{(4)}-\theta_{(2)})\, , \nn
\eea
where the $\star^{-1}$ operation is simply defined by 
\be
\label{starinv}
F \star^{-1} G = G \star F\, ,
\ee
where $F$ and $G$ represent superfields or products of superfields.

We will explain the effect of the $\star$-deformation by
considering the contribution of this graph to the $\la\bar\chi_1\chi^1\ra$
two-point function. For this contribution, we need to project the two
external legs in the supergraph onto the corresponding fermion
components. At external leg 1, this is achieved by acting with $\bar
d_{(1)1}$ and then setting $\theta_{(1)}^m=\bar\theta_{(1)m}=0$. This
yields
\be
\bar d_{(1)1}\,[d_{(1)}]^4[\bar d_{(1)}]^4\delta^8(\theta_{(1)}-
\theta_{(3)})\big|_{\theta_{(1)}=\bar\theta_{(1)}=0}
=\sqrt2p_-\,\bar\theta_{(3)1}\,\prod_{m=2}^4
[-1-\frac{p_-}{\sqrt2}\theta_{(3)}^m\bar\theta_{(3)m}]\, .
\ee
At external leg 2, the projection onto $\chi^1$
requires that we act with the operator $\bar d_{(2)2}\bar d_{(2)3}\bar
d_{(2)4}$ and then set $\theta_{(2)}^n=\bar\theta_{(2)n}=0$. This
computation gives
\bea
\bar d_{(2)2}\bar d_{(2)3}\bar d_{(2)4}\,[d_{(4)}]^4
\delta^8(\theta_{(4)}-\theta_{(2)})
\big|_{\theta_{(2)}=\bar\theta_{(2)}=0}=-\theta_{(4)}^1\,
\prod_{n=2}^4[1+\frac{p_-}{\sqrt2}\theta_{(4)}^n\bar\theta_{(4)n}]\, .
\eea
Having projected the two external legs onto the required fermionic
states, we now expand the piece between the $\star$-products as
\bea
[\bar d_{(3)}]^4[d_{(3)}]^4\delta^8(\theta_{(3)}-\theta_{(4)})
&\!\!=\!\!&\prod_{p=1}^4[-1+\sqrt2k_-\theta_{(3)}^p\bar\theta_{(4)p}
-\frac{1}{\sqrt2}k_-(\theta_{(3)}^p\bar\theta_{(3)p}+
\theta_{(4)}^p\bar\theta_{(4)p}) \nonumber \\
&&\hsp{0.65}-\frac{1}{2}k_-^2\theta_{(3)}^p\bar\theta_{(3)p}
\theta_{(4)}^p\bar\theta_{(4)p}]
\eea
The deformation will produce phase factors that we need to
identify. The phase factors due to the $\star_3$ in (\ref
{nptp}) arise from the following term
\be
\{\bar\theta_{(3)1}\star_3\prod_{p=1}^4\theta_{(3)}^p\}
\bar\theta_{(4)p}\, ,
\ee
while the phase factors from the $\star_4^{-1}$ are due to the term
\be
\prod_{p=1}^4\theta_{(3)}^p\{\bar\theta_{(4)p}\star_4^{-1}\theta_{(4)}^1\}\, .
\ee
These phase factors are easy to compute using table 2. Once the
$\star$-products have been evaluated, we are free to perform the
integral over $\theta_{(4)}$ with the help of the first
$\delta$-function in (\ref {nptp}). Since this sets
$\theta_{(3)}=\theta_{(4)}$, we will no longer explicitly write the
$(3)$ index in what follows. Thus, the contribution of (\ref {nptp})
to $\la\bar\chi_1\chi^1\ra$ is
\bea
\label{nptp2}
&&\int \dr^4\theta \,\dr^4\bar\theta\,\dr^4k\:\frac{(p-k)(\bar p-\bar k)}
{p_\mu^4p_-^2k_\nu^2{(p_--k_-)}^2{(p_\rho-k_\rho)}^2}\;(-2p_-^2)
[\theta^1\bar\theta_1][\theta^4\bar\theta_4] \nonumber \\
&&\times\,[\sqrt2\theta^2\bar\theta_2\{p_-k_-(\er^{i\pi\b}\er^{i\pi\b}-1)\}]
[\sqrt2\theta^3\bar\theta_3\{p_-k_-(\er^{-i\pi\b}\er^{-i\pi\b}-1)\}]\, .
\eea
This step illustrates the effect of the $\b$-deformation on non-planar
supergraphs. The shrinking of lines in non-planar graphs is
non-trivial due to the $\star(\ldots)\star^{-1}$ structure. This
structure limits our ability to move around $\star$'s (to free up a
$\delta$-function) and is responsible for the phases, from the two
$\star$'s, adding up and producing factors like
$\er^{-i\pi\b}\er^{-i\pi\b}$ in (\ref {nptp2}).

We now perform the remaining $\theta$-integration to obtain
\ba
&& \int \dr^4k \: \frac{(p-k)(\bar p-\bar k)}
{p_\mu^4p_-^2k_\nu^2{(p_--k_-)}^2 {(p_\rho-k_\rho)}^2}\nn \\
&&\hsp{1} \times \,(-2p_-^2)[p_-^2-2p_-k_-(\cos{2\pi\b}-1)-2k_-^2
(\cos{2\pi\b}-1)]\, .
\ea
Although the first term is logarithmic, the second and third are
linearly and quadratically divergent respectively. Thus the power
counting procedure of subsection~\ref {pcr} only offers a poor
upper bound on the superficial degree of divergence of this non-planar
supergraph, namely $D=2$. Thus the methods of section \ref {proof}
(which ensured the cancellation of the logarithmic divergences in
planar supergraphs) only prove the cancellation of quadratic
divergences in non-planar supergraphs.

\begin{figure}[!htb]
\begin{center}
\includegraphics[width=6cm]{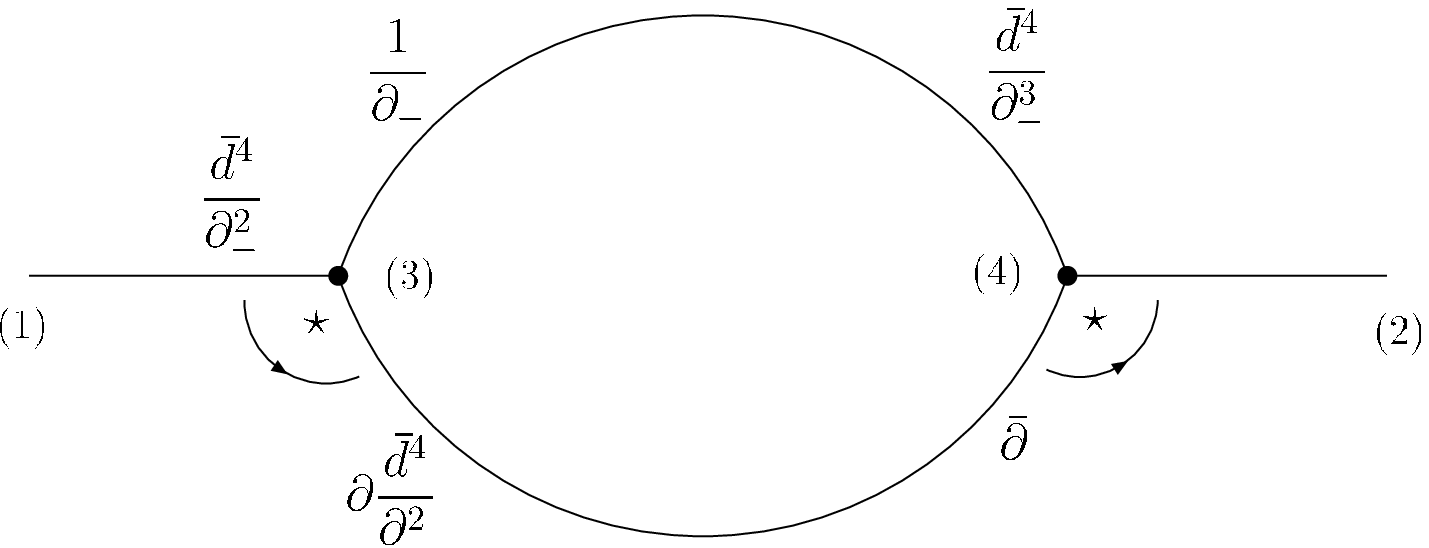}
\end{center}
\vsp{-0.4}
\caption{Planar contribution to the one-loop two-point function.}
\label{2pt-plan}
\end{figure}

\ndt
Having analyzed in detail the non-planar case we are in a position
to easily understand why planar supergraphs are much easier to
handle. The graph in figure 5 evaluates to
\bea
\label{ptp}
&&\int \dr^4\theta_{(3)}\,\dr^4\bar\theta_{(3)}\,\dr^4\theta_{(4)}
\,\dr^4\bar\theta_{(4)}\,\dr^4k\:\frac{(p-k)(\bar p-\bar k)}
{p_\mu^4p_-^2k_\nu^2{(p_--k_-)}^2{(p_\rho-k_\rho)}^2}\;
\delta^8(\theta_{(3)}-\theta_{(4)}) \\
&&\times \, [d_{(1)}]^4[\bar d_{(1)}]^4\delta^8(\theta_{(1)}
-\theta_{(3)})\,{\star_3}\,[\bar d_{(3)}]^4[d_{(3)}]^4
\delta^8(\theta_{(3)}-\theta_{(4)})\star_4[d_{(4)}]^4
\delta^8(\theta_{(4)}-\theta_{(2)})\, . \nn
\eea
We see immediately that the $\star$-structure differs from that in
(\ref {nptp}). This difference implies that the phase factors produced
in (\ref {nptp2}), instead of adding up now cancel. Once again, we
focus on the contribution of this two-point function to
$\la\bar\chi_1\chi^1\ra$. Proceeding in exactly the same manner
described so far, we find that this contribution is
\be
\frac{-2(p-k)(\bar p-\bar k)p_-^2}{p_\mu^4k_\nu^2{(p_--k_-)}^2
{(p_\rho-k_\rho)}^2}\, ,
\ee
which is logarithmically divergent. Since this graph has superficial
degree of divergence equal to zero, our treatment of it as described
in section \ref{proof} ensures that it is finite.

We remind the reader that in the planar limit, the one-loop
two-point function of the $\b$-deformed theory is identical to that in
$\calN=4$ Yang--Mills and has the correct asymptotic behavior at large
momentum.

\section{Quartic vertex contractions}
\label{details}

\subsection{Graphs involving two external legs}

The twenty-four contractions induced by the first quartic vertex in 
(\ref {ordergsq}) are
\bea
\label{fourv1abackup}
&&\hsp{-2}\Tr^\prime\left\{\fr{\parm}[\Delta_3,\parm\,
\Delta_4]_\star\,\fr{\parm}[\frac{\bar d^4}{\parm^2}\Delta_1,\frac{\bar d^4}
{\parm}\Delta_2]_\star+(1\leftrightarrow2)+(3\leftrightarrow4)
+(1\leftrightarrow2,3\leftrightarrow4) \right. \nn \\
&& \hsp{-1.45} + \fr{\parm}[\Delta_1,\parm\,\Delta_2]_\star\,\fr{\parm}
[\frac{\bar d^4}{\parm^2}\Delta_3,\frac{\bar d^4}{\parm}\Delta_4]_\star
+(1\leftrightarrow2)+(3\leftrightarrow4)
+(1\leftrightarrow2,3\leftrightarrow4) \nn \\
&& \hsp{-1.45} + \fr{\parm}[\Delta_3,\parm\,\Delta_1]_\star\,\fr{\parm}
[\frac{\bar d^4}{\parm^2}\Delta_4,\frac{\bar d^4}{\parm}\Delta_2]_\star
+(1\leftrightarrow2)+(3\leftrightarrow4)+(1\leftrightarrow2,
3\leftrightarrow4) \nn \\
&& \hsp{-1.45} + \fr{\parm}[\Delta_3,\parm\,\Delta_1]_\star\,\fr{\parm}
[\frac{\bar d^4}{\parm^2}\Delta_2,\frac{\bar d^4}{\parm}\Delta_4]_\star+
(1\leftrightarrow2)+(3\leftrightarrow4)+
(1\leftrightarrow2,3\leftrightarrow4) \nn \\
&& \hsp{-1.45} + \fr{\parm}[\Delta_2,\parm\,\Delta_3]_\star\,\fr{\parm}
[\frac{\bar d^4}{\parm^2}\Delta_4,\frac{\bar d^4}{\parm}\Delta_1]_\star
+(1\leftrightarrow2)+(3\leftrightarrow4)+(1\leftrightarrow2,
3\leftrightarrow4) \nn \\
&& \hsp{-1.45}\left.+\fr{\parm}[\Delta_1,\parm\,\Delta_3]_\star\,\fr{\parm}
[\frac{\bar d^4}{\parm^2}\Delta_2,\frac{\bar d^4}{\parm}\Delta_4]_\star
+(1\leftrightarrow2)+(3\leftrightarrow4)+(1\leftrightarrow2,
3\leftrightarrow4)\right\} \, .
\eea
The last line in the equation above was dealt with in section \ref
{qcont}. For the rest, the finiteness arguments are as follows. In
line $1$, we partially integrate out a factor $d^4$ (from the internal
propagator) of either internal leg to the external legs. Note that
these derivatives act in all possible ways on the two external legs
producing odd phase factors due to equation (\ref {newprop}). These
phase factors would be potentially dangerous if we were combining
terms to achieve finiteness. However, here the phase factors are
irrelevant because each individual term is itself finite. In line $2$,
both internal legs carry a factor ${\bar d}^4$. Starting from either
leg, this can be integrated out of the loop which becomes finite. The
numerator in line $3$ has a factor of $p_-$ and a factor of $q_-$ both
of which do not contribute to the integral. Lines $4$ and $5$ both
involve a factor of $p_-$ in the numerator. These factors ensure that
the integrals resulting from lines $3$, $4$ and $5$ are finite.

As explained in section \ref {qcont}, the twenty-four contractions from
the second quartic vertex in (\ref {ordergsq}) reduce to twelve
terms. Four of these twelve terms can be easily shown to be finite
using the manipulations described in the main text. This leaves eight
terms
\ba
&& \Tr^\prime\left\{[\Delta_1,\frac{{\bar d}^4}{\parm^2}\Delta_2]_\star
[\Delta_3,\frac{{\bar d}^4}{\parm^2}\Delta_4]_\star+[\Delta_1,
\frac{{\bar d}^4}{\parm^2}\Delta_2]_\star[\Delta_4,
\frac{{\bar d}^4}{\parm^2}\Delta_3]_\star \right. \nn \\
&& \hsp{0.55}+ [\Delta_2,\frac{{\bar d}^4}{\parm^2}\Delta_1]_\star
[\Delta_3,\frac{{\bar d}^4}{\parm^2}\Delta_4]_\star+[\Delta_2,
\frac{{\bar d}^4}{\parm^2}\Delta_1]_\star[\Delta_4,
\frac{{\bar d}^4}{\parm^2}\Delta_3]_\star  \nn \\
&& \hsp{0.55}+ [\Delta_1,\frac{{\bar d}^4}{\parm^2}\Delta_3]_\star
[\Delta_4,\frac{{\bar d}^4}{\parm^2}\Delta_2]_\star+[\Delta_1,
\frac{{\bar d}^4}{\parm^2}\Delta_4]_\star[\Delta_3,
\frac{{\bar d}^4}{\parm^2}\Delta_2]_\star  \nn \\
&& \hsp{0.5}\left.+ [\Delta_3,\frac{{\bar d}^4}{\parm^2}
\Delta_1]_\star[\Delta_2,\frac{{\bar d}^4}{\parm^2}\Delta_4]_\star
+[\Delta_4,\frac{{\bar d}^4}{\parm^2}\Delta_1]_\star[\Delta_2,
\frac{{\bar d}^4}{\parm^2}\Delta_3]_\star \right\}  . 
\label{fourv2}
\ea
The last two lines in the above equation were dealt with in section
\ref {qcont}. Here, we briefly explain why the first two lines are
finite. In the first term, we integrate the ${\bar d}^4$ away from
$\Delta_4$. If even one $\bar d$ is integrated to external leg $1$,
the term becomes finite. So we focus on the case where the four ${\bar
d}$'s move to the other internal leg. This reads
\bea
[\Delta_1,\frac{{\bar d}^4}{\parm^2}\Delta_2]_\star
[{\bar d}^4\Delta_3,\frac{1}{\parm^2}\Delta_4]_\star\, ,
\eea
In terms of momenta, the first line of (\ref {fourv2}) is now
\bea
\label{t12}
{\biggl (}\fr{q_-^2}\fr{l_-^2}-\fr{q_-^2}
\fr{k_-^2}{\biggr )}\;[\Delta_1,{\bar d}^4\Delta_2]_\star
[{\bar d}^4\Delta_3,\Delta_4]_\star\, .
\eea
Momentum conservation gives
\bea
k=p+q-l \, ,\qquad k=-l\;\;{\mbox {for}}\;\;l\,{\gg}\,p,q\, ,
\eea
which implies that the divergent part of (\ref {t12}) vanishes. The
proof of finiteness for the second line in (\ref {fourv2}) follows
from similar arguments.


\begin{thebibliography}{Ref}

\bibitem{N4}{L. Brink, J. Schwarz and J. Scherk, {\it Nucl. Phys.} 
{\bf B 121} (1977) 77. \\
F. Gliozzi, J. Scherk and D. Olive, {\it Nucl. Phys.} 
{\bf B 122} (1977) 256.}
\bibitem{finN4}{L.~V.~Avdeev, O.~V.~Tarasov and A.~A.~Vladimirov, 
{\it Phys. Lett.} {\bf B 96} (1980) 94. \\
M.~T.~Grisaru, M.~Rocek and W.~Siegel, {\it Phys. Rev. Lett.}  
{\bf 45} (1980) 1063. \\
M.~F.~Sohnius and P.~C.~West, {\it Phys. Lett.} {\bf B 100} 
(1981) 245. \\
W.~E.~Caswell and D.~Zanon, {\it Nucl. Phys.} {\bf B 182} (1981)
125. \\ 
P.~S.~Howe, K.~S.~Stelle and P.~K.~Townsend, {\it Nucl. Phys.} 
{\bf B 236} (1984) 125.}
\bibitem{SM}{S. Mandelstam, {\it Nucl. Phys.} {\bf B 213} (1983) 149.}
\bibitem{BLN2}{L.~Brink, O.~Lindgren and B.~E.~W.~Nilsson, 
{\it Phys. Lett.} {\bf B 123} (1983) 323.}
\bibitem{oldfinite}{P.~C.~West, {\it Phys. Lett.} {\bf B 137} (1984)
371. \\
A.~Parkes and P.~C.~West, {\it Phys. Lett.} {\bf B 138} (1984) 99. \\
D.~R.~T.~Jones and L.~Mezincescu,{\it Phys. Lett.} {\bf B 138} (1984)
293. \\ 
S.~Hamidi, J.~Patera and J.~H.~Schwarz, {\it Phys. Lett.}  {\bf B 141}
(1984) 349. \\
S.~Hamidi and J.~H.~Schwarz, {\it Phys. Lett.} {\bf B 147} (1984)
301. \\ 
W.~Lucha and H.~Neufeld, {\it Phys. Lett.} {\bf B 174} (1986) 186; 
{\it Phys. Rev.} {\bf D 34} (1986) 1089. \\
D.~R.~T.~Jones, {\it Nucl. Phys.} {\bf B 277} (1986) 153. \\
A.~V.~Ermushev, D.~I.~Kazakov and O.~V.~Tarasov,
{\it Nucl. Phys.} {\bf B 281} (1987) 72. \\
X.~-D.~Jiang and X.~-J.~Zhou, {\it Phys. Rev.} {\bf D 42} (1990)
2109; {\it Phys. Lett.} {\bf B 197} (1987) 156; {\it Phys. Lett.}
{\bf B 216} (1989) 160. \\ 
D.~I.~Kazakov, {\it Mod. Phys. Lett.} {\bf A 2} (1987) 663. \\
O.~Piguet and K.~Sibold, {\it Phys. Lett.} {\bf B 177} (1986) 373; 
{\it Int. J. Mod. Phys.}  {\bf A 1} (1986) 913. \\
C.~Lucchesi, O.~Piguet and K.~Sibold, {\it Phys. Lett.} {\bf B 201}
(1988) 241. \\
N.~Marcus and A.~Sagnotti, {\it Nucl. Phys.} {\bf B 256} (1985) 77.
}
\bibitem{LS}{R. G. Leigh and M. J. Strassler, {\it Nucl. Phys.} 
{\bf B 447} (1995) 95 [\hepth{9503121}].}
\bibitem{adscft}{J. M. Maldacena, {\it Adv. Theor. Math. Phys.} 
{\bf 2} (1998) 231 [\hepth{9711200}]. \\
S. S. Gubser, I. R. Klebanov and A. M. Polyakov, 
{\it Phys. Lett.} {\bf B 428} (1998) 105 [\hepth{9802109}]. \\
E. Witten, {\it Adv. Theor. Math. Phys.} 
{\bf 2} (1998) 253 [\hepth{9802150}].}
\bibitem{bl}{D.~Berenstein and R.~G.~Leigh, {\it JHEP} {\bf 0001}
(2000) 038 [\hepth{0001055}].}
\bibitem{AKY}{O.~Aharony, B.~Kol and S.~Yankielowicz,
{\it JHEP} {\bf 0206} (2002) 039 [\hepth{0205090}].}
\bibitem{LM}{O. Lunin and J. M. Maldacena, {\it JHEP} {\bf 0505} 
(2005) 033 [\hepth{0502086}].}
\bibitem{fg}{D.~Z.~Freedman and U.~Gursoy, {\it JHEP} {\bf 0511} 
(2005) 042 [\hepth{0506128}].}
\bibitem{rss}{G.~C.~Rossi, E.~Sokatchev and Ya.~S.~Stanev,
{\it Nucl. Phys.} {\bf B 729} (2005) 581 [\hepth{0507113}].} \\
{G.~C.~Rossi, E.~Sokatchev and Ya.~S.~Stanev, 
\hepth{0606284}.}
\bibitem{milan}{S.~Penati, A.~Santambrogio and D.~Zanon,
{\it  JHEP} {\bf 0510} (2005) 023 [\hepth{0506150}]. \\
A. Mauri, S. Penati, A. Santambrogio and D. Zanon, 
{\it JHEP} {\bf 0511} (2005) 024 [\hepth{0507282}]. \\
A.~Mauri, S.~Penati, M.~Pirrone, A.~Santambrogio and D.~Zanon, 
{\it JHEP} {\bf 0608} (2006) 072 [\hepth{0605145}].}
\bibitem{vk}{V.~V.~Khoze, {\it JHEP} {\bf 0602} (2006) 040
[\hepth{0512194}].}
\bibitem{BLN1}{L. Brink, O. Lindgren and B. E. W. Nilsson, 
{\it Nucl. Phys.} {\bf B 212} (1983) 401.}
\bibitem{BBB2}{A. K. H. Bengtsson, I. Bengtsson and L. Brink, 
{\it Nucl. Phys.} {\bf B 227} (1983) 41.}
\bibitem{SA1}{S. Ananth, Ph.D. Thesis, ISBN: 0542303965 (2005).}
\bibitem{ABR1}{S. Ananth, L. Brink and P. Ramond, {\it JHEP} 
{\bf 0407} (2004) 082 [\hepth{0405150}].}
\bibitem{ABKR}{S. Ananth, L. Brink, S. Kim and P. Ramond, 
{\it Nucl. Phys. } {\bf B 722} (2005) 166 [\hepth{0505234}].}
\bibitem{BT}{L.~Brink and A.~Tollsten, {\it Nucl. Phys.}  {\bf B 249} 
(1985) 244.}
\bibitem{SA2}{S.~Ananth, {\it JHEP} {\bf 0512} (2005) 010
[\hepth{0510064}].}
\bibitem{noncomm}{A.~Connes, M.~R.~Douglas and A.~S.~Schwarz,
{\it JHEP} {\bf 9802} (1998) 003 [\hepth{9711162}]. \\
V. Schomerus, {\it JHEP} {\bf 9906} (1999) 030 [\hepth{9903205}]. \\
N. Seiberg and E. Witten, {\it JHEP} {\bf 9909} (1999) 032 
[\hepth{9908142}]. \\
N.~Seiberg, {\it JHEP} {\bf 0306} (2003) 010 [\hepth{0305248}].}
\bibitem{GRS}{M. Grisaru, M. Rocek and W. Siegel, 
{\it Nucl. Phys.} {\bf B 159} (1979) 429.}
\bibitem{SW}{S. Weinberg, {\it Phys. Rev.} {\bf 118} (1960) 838.}
\bibitem{BPP}{S. J. Brodsky, H. C. Pauli and S. S. Pinsky, 
{\it Phys. Rept.} {\bf 301} (1998) 299 [\hepph{9705477}]. \\
T.~Heinzl, {\it Lect. Notes Phys.} {\bf 572} (2001) 55
[\hepth{0008096}].}
\bibitem{Z}{W.~Zimmermann, {\it Comm. Math. Phys.} {\bf 11} (1968) 1.}
\bibitem{vvk}{G.~Georgiou and V.~V.~Khoze, {\it JHEP} {\bf 0604}
(2006) 049 [\hepth{0602141}]. \\
C.~Durnford, G.~Georgiou and V.~V.~Khoze, \hepth{0606111}. \\
G.~Georgiou, V.~V.~Khoze and S.~Kovacs, in preparation.}
\bibitem{instN4}{M.~Bianchi, M.~B.~Green, S.~Kovacs and G.~C.~Rossi,
{\it JHEP} {\bf 9808} (1998) 013 [\hepth{9807033}]. \\
N.~Dorey, T.~J.~Hollowood, V.~V.~Khoze, M.~P.~Mattis and S.~Vandoren,
{\it Nucl. Phys.}  {\bf B 552} (1999) 88 [\hepth{9901128}].}
\bibitem{dhk}{N.~Dorey, T.~J.~Hollowood and S.~P.~Kumar,
{\it JHEP} {\bf 0212} (2002) 003 [\hepth{0210239}].}
\bibitem{SF}{S.~Frolov, {\it JHEP} {\bf 0505} (2005) 069
[\hepth{0503201}].}
\bibitem{np}{V.~Niarchos and N.~Prezas, {\it JHEP} {\bf 0306} (2003) 015
[\hepth{0212111}].}
\bibitem{bjl}{D.~Berenstein, V.~Jejjala and R.~G.~Leigh,
{\it Nucl. Phys.} {\bf B 589} (2000) 196 [\hepth{0005087}].}
\bibitem{MZ}{J. A. Minahan and K. Zarembo, {\it JHEP} {\bf 0303} 
(2003) 013 [\hepth{0212208}].} \\
{N. Beisert, C. Kristjansen and M. Staudacher, {\it Nucl. Phys.} 
{\bf B 664} (2003) 131 [\hepth{0303060}].} \\
{N. Beisert and M. Staudacher, {\it Nucl. Phys.} {\bf B 670} (2003) 439 
[\hepth{0307042}].}
\end{thebibliography}
\end{document}